\documentclass{article}

\usepackage{arxiv}

\usepackage[utf8]{inputenc} 
\usepackage[T1]{fontenc}    
\usepackage{hyperref}       
\usepackage{url}            
\usepackage{booktabs}       
\usepackage{amsfonts}       
\usepackage{nicefrac}       
\usepackage{microtype}      
\usepackage{lipsum}     
\usepackage{graphicx}
\usepackage{doi}

\usepackage{amsmath}
\usepackage{amssymb}
\usepackage{latexsym}
\usepackage{graphicx}
\usepackage{natbib}
\usepackage{color}
\bibpunct{(}{)}{;}{a}{}{,}

\usepackage{amstext}

\newcommand{\arcsec}{\ensuremath{^{\prime\prime}}}
\newcommand{\apj}{ApJ}
\newcommand{\aap}{A\&A}
\newcommand{\apjl}{ApJL}
\newcommand{\apjs}{ApJS}
\newcommand{\nat}{Nature}
\newcommand{\solphys}{Sol. Phys.}
\newcommand{\ssr}{Space Sci. Rev.}

\DeclareRobustCommand{\okina}{%
  \raisebox{\dimexpr\fontcharht\font`f-\height}{%
    \scalebox{0.8}{`}%
  }%
}

\title{High-frequency coronal Alfv\'enic waves observed with DKIST/Cryo-NIRSP}

\author{ \href{https://orcid.org/0000-0001-5678-9002}{\includegraphics[scale=0.06]{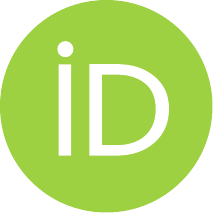}\hspace{1mm}Richard J. Morton}\\
Department of Maths Physics and Electrical Engineering \\
Northumbria University, UK \\
\texttt{richard.morton@northumbria.ac.uk}\\
\And
\href{https://orcid.org/0000-0003-0583-0516}{\includegraphics[scale=0.06]{orcid.pdf}\hspace{1mm}Momchil Molnar}\\
Southwest Research Institute\\ Boulder\\ CO 80302, USA \\
\And
\href{https://orcid.org/0000-0002-3699-3134}{\includegraphics[scale=0.06]{orcid.pdf}\hspace{1mm}Steven R. Cranmer}\\
Department of Astrophysical and Planetary Sciences\\
Laboratory for Atmospheric and Space Physics\\
University of Colorado\\ Boulder\\ CO 80309, USA \\
\And
\href{https://orcid.org/0000-0002-7451-9804}{\includegraphics[scale=0.06]{orcid.pdf}\hspace{1mm}Thomas A. Schad}\\
National Solar Observatory\\ 22 \okina\={O}hi\okina a K\={u} Street\\ Pukalani\\ HI 96768, USA}

\begin{document}
\maketitle

\begin{abstract}
The presence and nature of low-frequency (0.1-10~mHz) Alfv\'enic waves in the corona has been established over the last decade, with many of these results coming from coronagraphic observations of the infrared Fe XIII line. The Cryo-NIRSP instrument situated at DKIST has recently begun acquiring science quality data of the same Fe XIII line, with at least a factor of 9 improvement in spatial resolution, a factor 30 increase in temporal resolution and an increase in signal-to-noise, when compared to the majority of previously available data. Here we present an analysis of 1~s cadence sit-and-stare data from Cryo-NIRSP, examining the Doppler velocity fluctuations associated with the Fe XIII 1074~nm coronal line. We are able to confirm previous results of Alfv\'enic waves in the corona as well as explore a new frequency regime. The data reveals that the power law behaviour of the Doppler velocity power spectrum extends to higher frequencies. This result appears to challenge some models of photospheric-driven Alfv\'enic waves that predict a lack of high frequency wave power in the corona due to strong chromospheric damping. Moreover, the high-frequency waves do not transport as much energy as their low-frequency counterparts, with less time-averaged energy per frequency interval. We are also able to confirm the incompressible nature of the fluctuations with little coherence between the line amplitude and Doppler velocity time-series.
\end{abstract}


\section{Introduction} \label{sec:intro}
The potential role of Alfv\'enic waves in being a key transporter of energy through the Sun's atmosphere is well known. Many analytic and numerical models have demonstrated the capability for Alfv\'enic waves to effectively move energy from the photosphere out to the corona and solar wind, depositing the energy to heat plasma and add momentum to outflows \citep[e.g.,][]{SUZINU2005,CRAVAN2005,CRAetal2007,VANBALLetal2011,van_der_Holst_2014,Shoda_2018}. 

The generally accepted theory for Alfv\'enic wave propagation from the photosphere to the corona contains several barriers for getting Alfv\'enic waves into the corona. The waves are considered to be largely excited by the convective motions in the photosphere, which can twist and buffet magnetic field lines \citep[e.g.,][]{CRAVAN2005,FEDetal2011b,VIGetal2012,Shelyag_2013}. There is substantial observational evidence for photospheric flux concentrations being displaced transversally \citep[e.g.,][]{VANBALLetal1998,Hasan_2000,NISetal2003,CHITetal2012,STAetal2013,STAetal2014} and being subject to vortex or swirling flows \citep{BONetal2008,BALetal2010,WEDetal2012,MORetal2013,2018ApJ...869..169G}. The presence of waves in the chromosphere is well established, supporting the theory of photospheric excitation. Many observations using high-resolution imaging and spectroscopic data demonstrate that much of the chromospheric fine structure (i.e., fibrils and spicules) supports transverse non-axisymmetric \citep[kink, e.g.,][]{DEPetal2007,OKADEP2011,MORetal2012c,PERetal2012,MORetal2013,MOOetal2017,MORetal2021,Chae_2021} and torsional Alfv\'enic modes \citep{De_Pontieu_2012,De_Pontieu_2014}. 

In order to reach the corona, the waves have to navigate the transition region. Due to the significant gradient in Alfv\'en speed that occurs there, low-frequency Alfv\'enic waves can be substantially reflected \citep[e.g.,][]{Hollweg_1981,CRAVAN2005,2017ApJ...840...20S, Pelouze_2023}. For those that reach the corona, the low-frequency waves have been thought to be difficult to dissipate (e.g., via phase mixing) without perpendicular gradients in the plasma. Hence, high-frequency Alfv\'enic waves have been suggested to be important in powering the corona \citep[e.g.,][]{axford1992,MarchTu1997}. However, the high frequency waves excited in the lower atmosphere are potentially subject to significant damping from ion-neutral effects in the photosphere and chromosphere \citep[e.g.,][]{2011A&A...534A..93Z,Soler_2015b,2017ApJ...840...20S,2019ApJ...871....3S}. It should be said that turbulence models of Alfv\'enic wave propagation negate the issues around liberating energy from the low frequency waves, as the energy from these motions is cascaded to smaller scales where it can be effectively dissipated \citep[e.g.,][]{MATetal1999}.

\medskip

Despite these difficulties, we know that the Alfv\'enic waves are present in the corona. The first direct identification of ubiquitous Alfv\'enic waves in the corona was provided by \cite{TOMetal2007}. The characteristic signals of the waves were interpreted from measurements of the 1074~nm Fe XIII emission line in the infrared. The observations were made using the Coronal MultiChannel Polarimeter \citep[CoMP - ][]{TOMetal2008} instrument, which is an imaging spectropolarimeter. CoMP was able to reveal the presence of fast propagating Doppler velocity signals, whose propagation speed significantly exceeded the local coronal sound speed and were found to have no counterpart oscillations in intensity. Earlier coronagraphic studies of the Fe XIV 5303~{\AA} line had also found the presence of Doppler velocity oscillations with no intensity counterpart \citep{Tsubaki_1977}. Since the first identification of coronal Alfv\'enic waves, CoMP has provided routine measurements of the 1074~nm line, and subsequent studies have reconfirmed the veracity of the original results \citep{THRetal2013,MORetal2015,MORetal2016,MORetal2019,YANGetal2020}. The studies also identified various properties of the coronal Alfv\'enic waves, namely that they had a distinct power spectrum between 0.1-10~mHz that was comprised of a power law with an enhancement of power around 4~mHz \citep{MORetal2016,MORetal2019}. The presence of the power enhancement has been put forward \citep[e.g.,][]{MORetal2019} as support for a theory of additional Alfv\'enic wave energy injection into the corona  through the mode conversion of \textit{p}-modes \citep[e.g.,][]{CALGOO2008, Miriyala_2025}. Some remanent of the enhancement may also persist as the waves propagate outwards, with Parker Solar Probe data indicating that fluctuation energy is concentrated around time-scales of 3-8~mHz in the outer corona \citep[at $\sim10$~$R_\odot$][]{Huang__2024}.

\medskip 

The previous observations of Alfv\'enic waves in the low corona have been limited in various ways. The CoMP data was generally restricted to cadences of 30~s and spatial resolutions of $\sim4.5^{\arcsec}$ ($\sim3500$~km). The low spatial resolution led to the integration of the emission line profiles over a number of neighbouring fine-scale coronal structures, whose cross-sectional widths measured in imaging data range from 100-2000~km  \citep{Antolin_2012,BROetal2013,PETetal2013,MorCun2023}. However simulations suggest an almost continuous spectrum of scales should be present in the corona \citep[e.g.,][]{Malnushenko2022}. 

The coronal Alfv\'enic waves have also been observed on individual fine-structures with the Solar Dynamics Observatory (SDO) \citep{MCIetal2011,THUetal2014,MORetal2019,Weberg_2020}, which show transverse displacements of the structures in the plane-of-sky. Hence both the Doppler velocities fluctuations observed with CoMP and the physical displacements are associated with the Alfv\'enic kink mode \citep{GOOetal2012}. There is an apparent discrepancy between the results, with the estimated Doppler velocity having apparently low amplitude fluctuations, with $v_{rms}\le 1$~km/s, while the observed transverse displacements had velocity amplitudes of $15-30$~km/s. However, the comparatively lower Doppler amplitudes may be explained by the integration of many out-of-phase oscillating structures along the line of sight \citep{DEMPAS2012,Pant_2019,2022PhDT........10G}.

The relatively low cadence of the CoMP data is restrictive, as it can only resolve frequencies less than $\sim10$~mHz. Waves with high frequencies are also difficult to observe with SDO as the amplitude of the transverse displacement decreases with frequency \citep{MORetal2019,Weberg_2020}, hence they are unable to be resolved in images. Therefore, previous wave observations have been restricted to a relatively low frequency range and our knowledge of the high-frequency waves is still substantially lacking. Coincidentally, results from theoretical models of Alfv\'en wave damping in partially ionised plasmas suggest that around 10~mHz there is a steep decline in transmission of wave energy into the corona due to wave damping \citep{2019ApJ...871....3S}. Hence, it is interesting to be able to determine the high-frequency behaviour of the coronal Alfv\'enic waves. The Cryogenic Near-Infrared Spectropolarimeter \citep[Cryo-NIRSP -][]{Fehlmann_2023} instrument has the potential to provide unique insights into high-frequency dynamics in the corona, being able to achieve high-cadence ($\sim1$~s) observations of the Fe XIII emission line. The first spectra were reported by \cite{Schad_2023} and showed exquisite detail. Here we provide the first investigation of the coronal Doppler velocities fluctuations with Cryo-NIRSP, showing the potential for investigation of high-frequency waves.

\section{Observations}
\begin{figure*}
    \centering
    \includegraphics[trim=60 0 60 10, clip, scale=0.6]{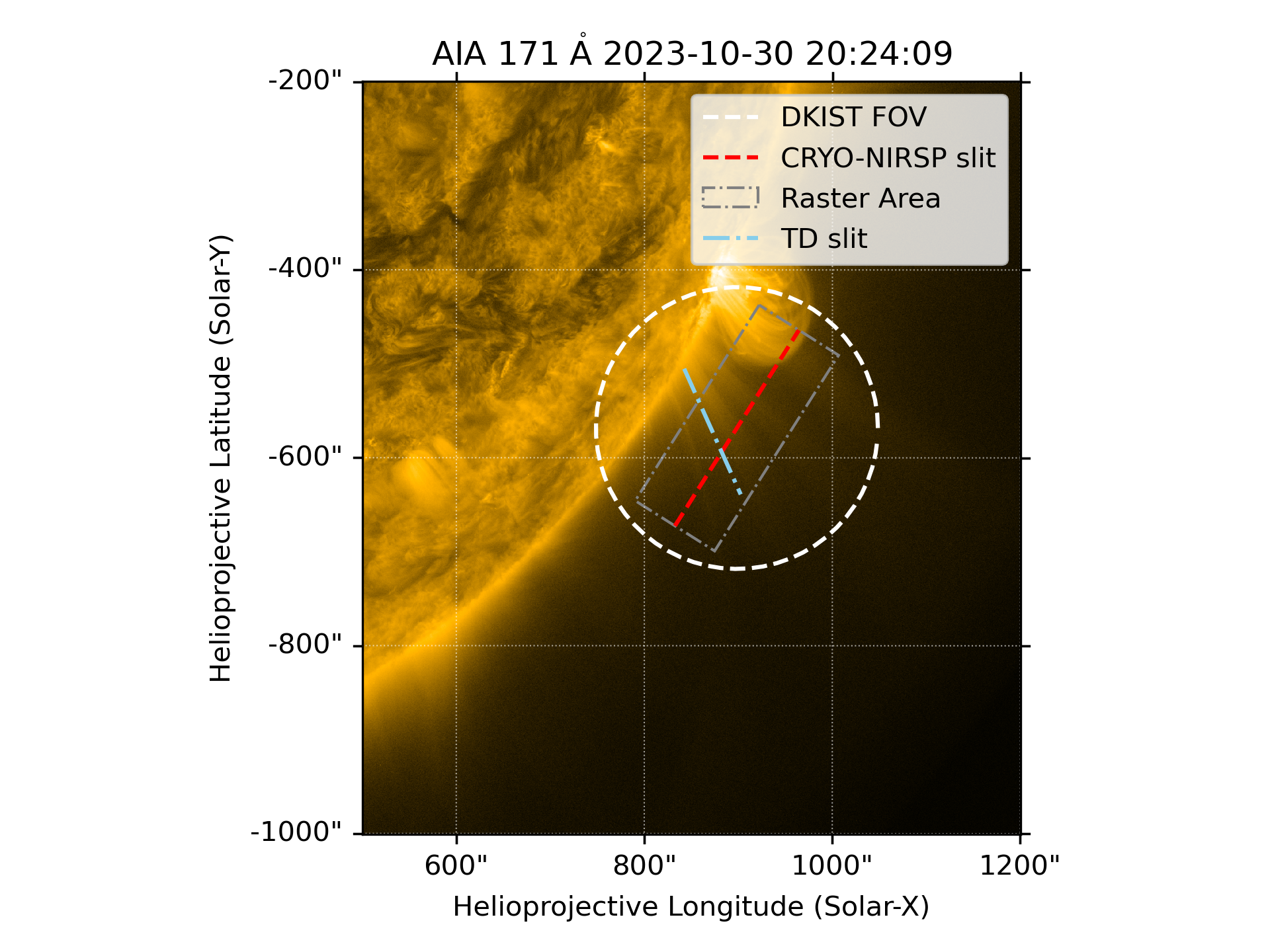}
    \includegraphics[trim=50 0 60 10, clip, scale=0.6]{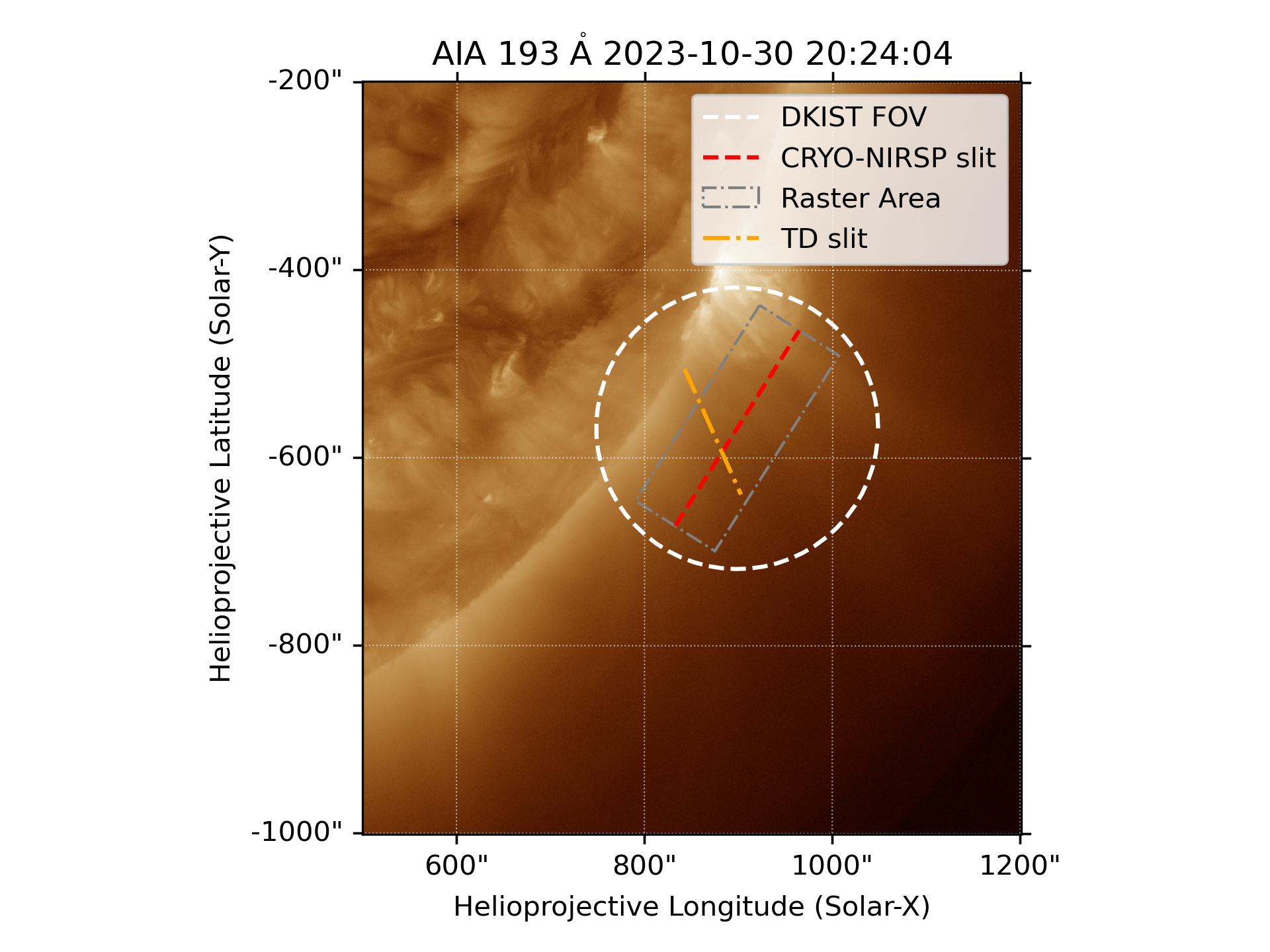}
    \includegraphics[trim=45 0 60 25, clip, scale=0.6]{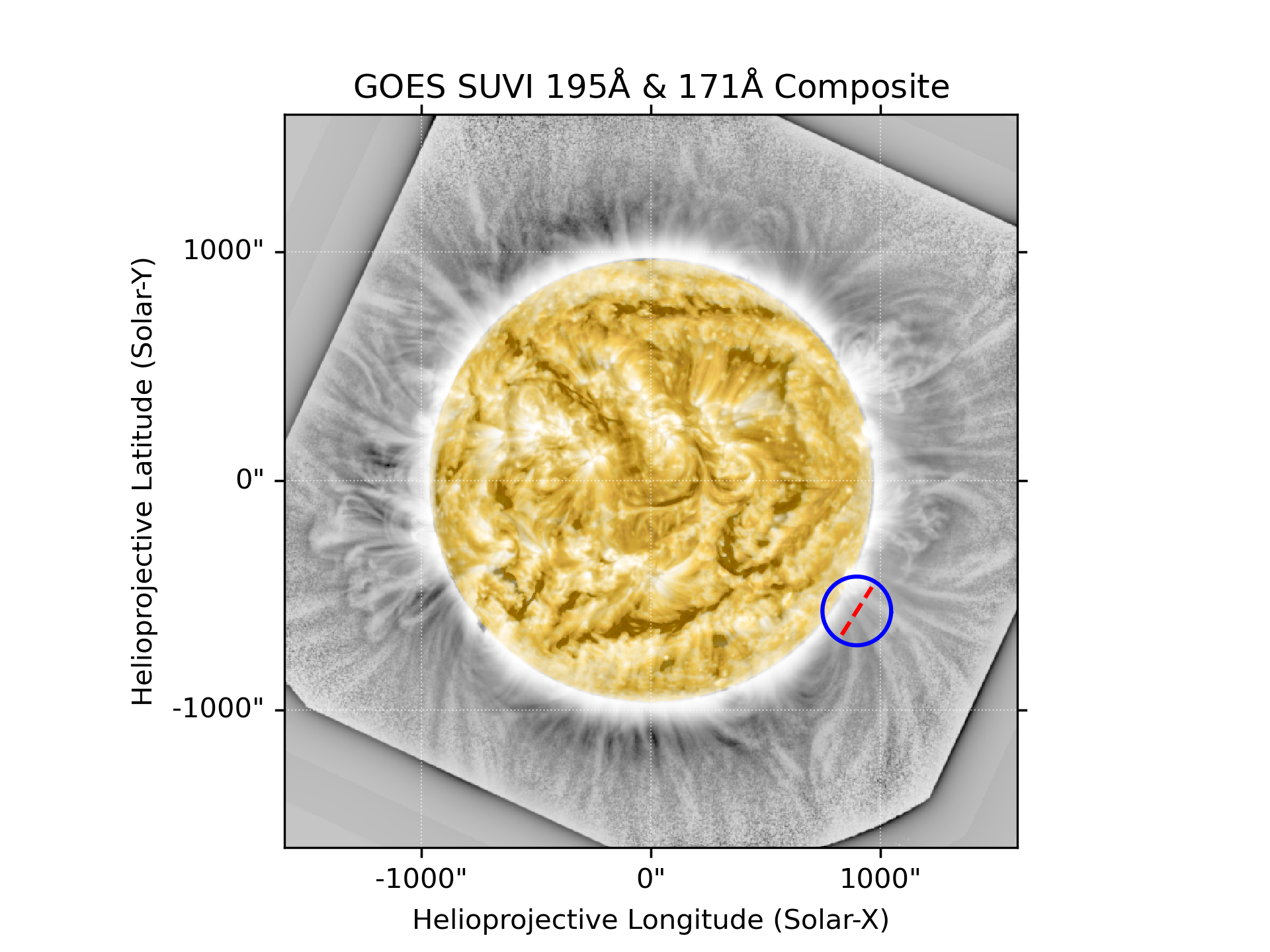}
    \includegraphics[trim=50 0 60 25, clip, scale=0.6]{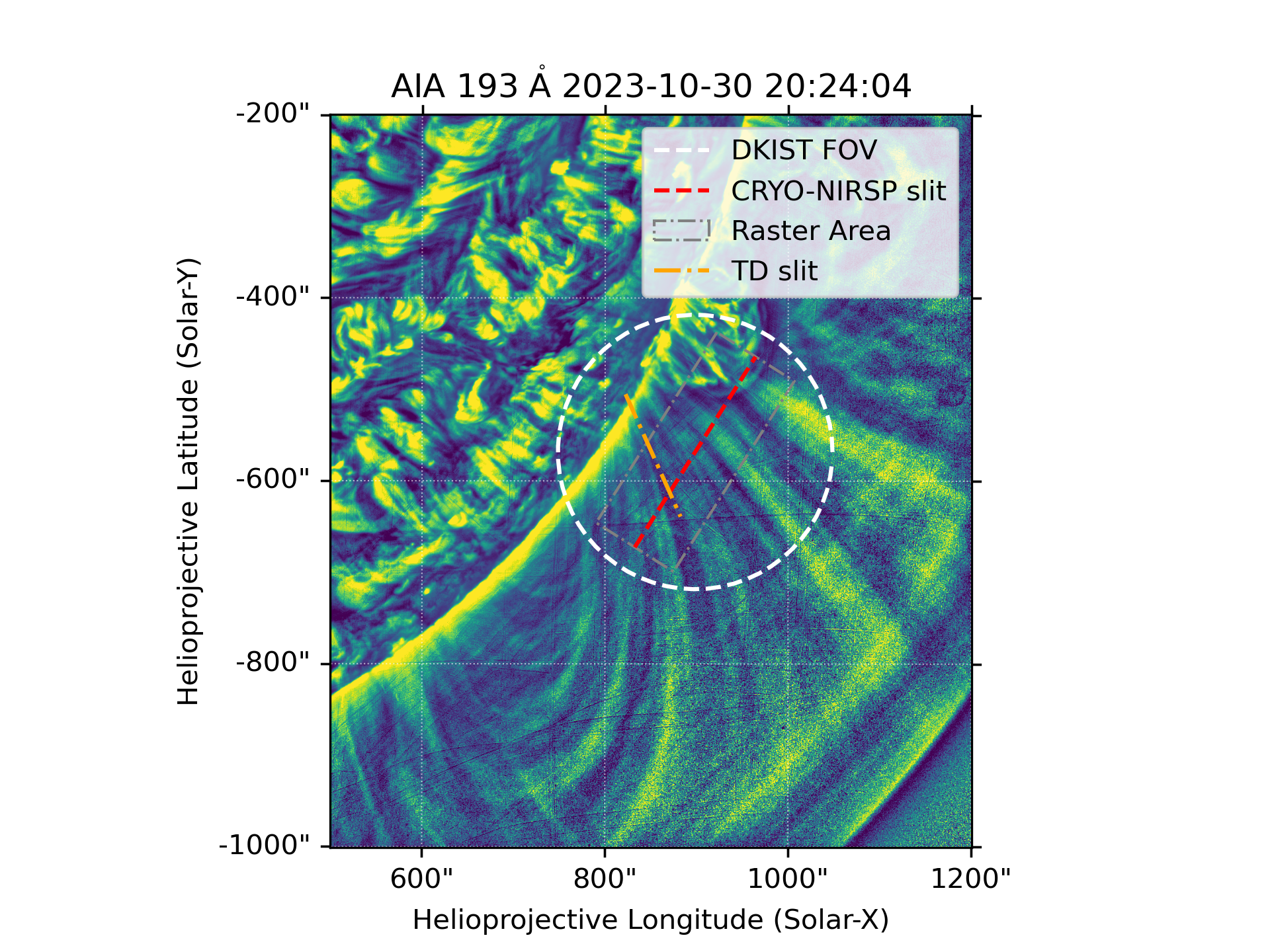}
    \caption{Context observations in the EUV. The upper left and right panels are images from SDO/AIA taken in the 171~{\AA} and 193~{\AA} passbands respectively. The bottom left panel shows a composite SUVI image processed with multi-Gaussian normalisation to bring out the details of the coronal structure. The data at the limb is in the 195~{\AA} passband and the disk image is from 171~{\AA} passband. The bottom right is the SDO/AIA 193~{\AA} passband image filtered with multi-Gaussian normalisation \citep{MORDRU2014}. The circular DKIST field-of-view, the Cryo-NIRSP raster region and the Cryo-NIRSP slit position for the sit-and-stare observations are over-plotted on the images. The SDO/AIA images also indicate the location of data extracted for time-distance analysis (TD slit). }\label{fig:overview}
\end{figure*}

We present here off-limb observations of the Fe~XIII 1074~nm coronal line that were obtained by the Cryo-NIRSP instrument on 30 October 2023 (20:24~UTC until 23:00~UTC). For context, in Figure~\ref{fig:overview} we display an EUV image from SDO \citep{PESetal2012} Atmospheric Imaging Assembly \citep[][]{LEMetal2012} and also an image from the GOES Solar Ultraviolet Imager \citep[SUVI;][]{Darnel_2022} with the DKIST field of view over-plotted. The Cryo-NIRSP observations consisted of two modes of operation. The first was a raster-scan of the target region, which was centred on $(X,Y) =(898,-568)$ (a height of 0.1~$R_\odot$ above the limb) and scanned a region of 100$^{\arcsec}$ in length with a 0.5$^{\arcsec}$ step between samples. The second operation mode was a sit-and-stare observation, with the slit also centred $(X,Y) =(898,-568)$. In both cases, the slit has a 0.12$^{\arcsec}$ per pixel sampling along the slit and a 0.5$^{\arcsec}$ width. We assess the pointing stability of the data by examining the motions of the fine structure in the sit-and-stare observation, and it appears to be less than 0.1$^{\arcsec}$. Hence, the stability of the pointing is better than the estimated seeing resolution along the slit ($\sim0.6^{\arcsec}$ - see Sections~\ref{sec:coher} and \ref{sec:coher_discuss}).

\begin{figure}
    \centering
    \includegraphics[trim=0 0 0 0, clip, scale=0.45]{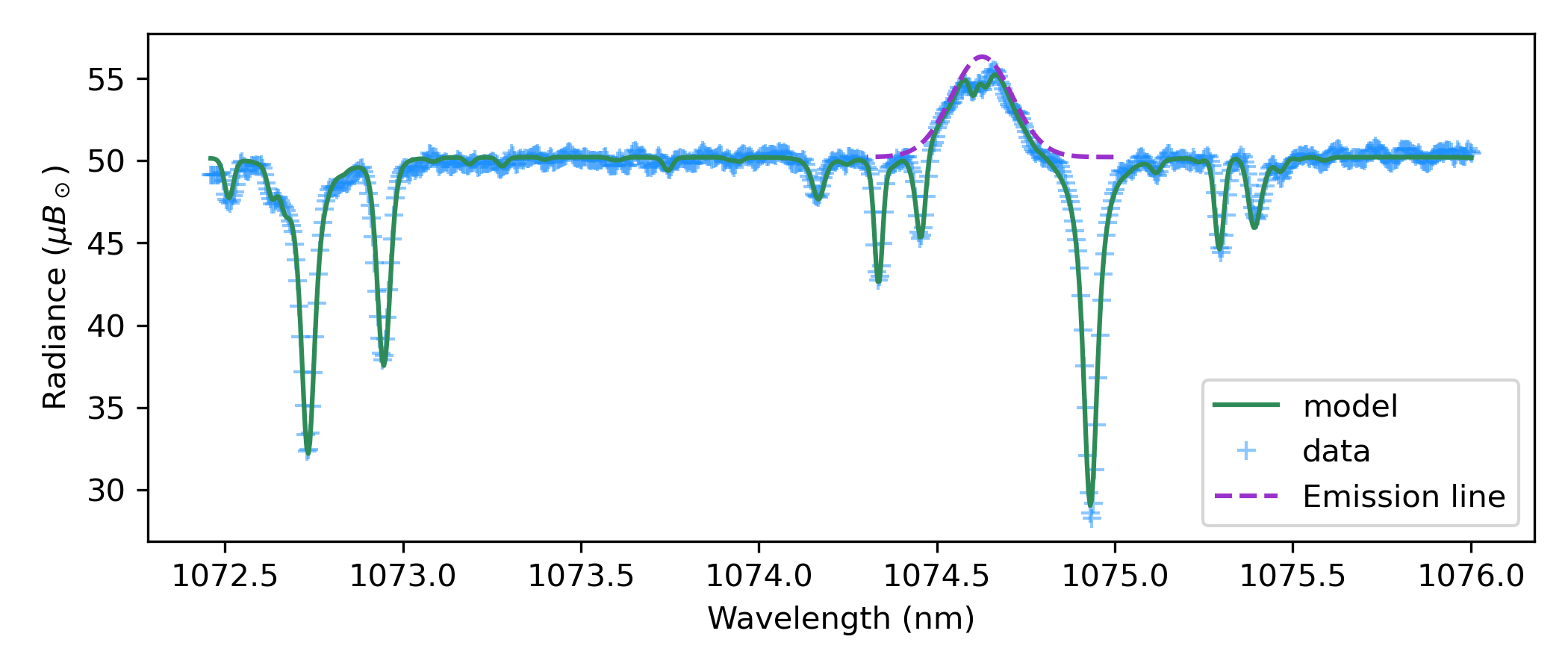} \\
    \includegraphics[trim=10 70 40 80, clip, scale=0.6]{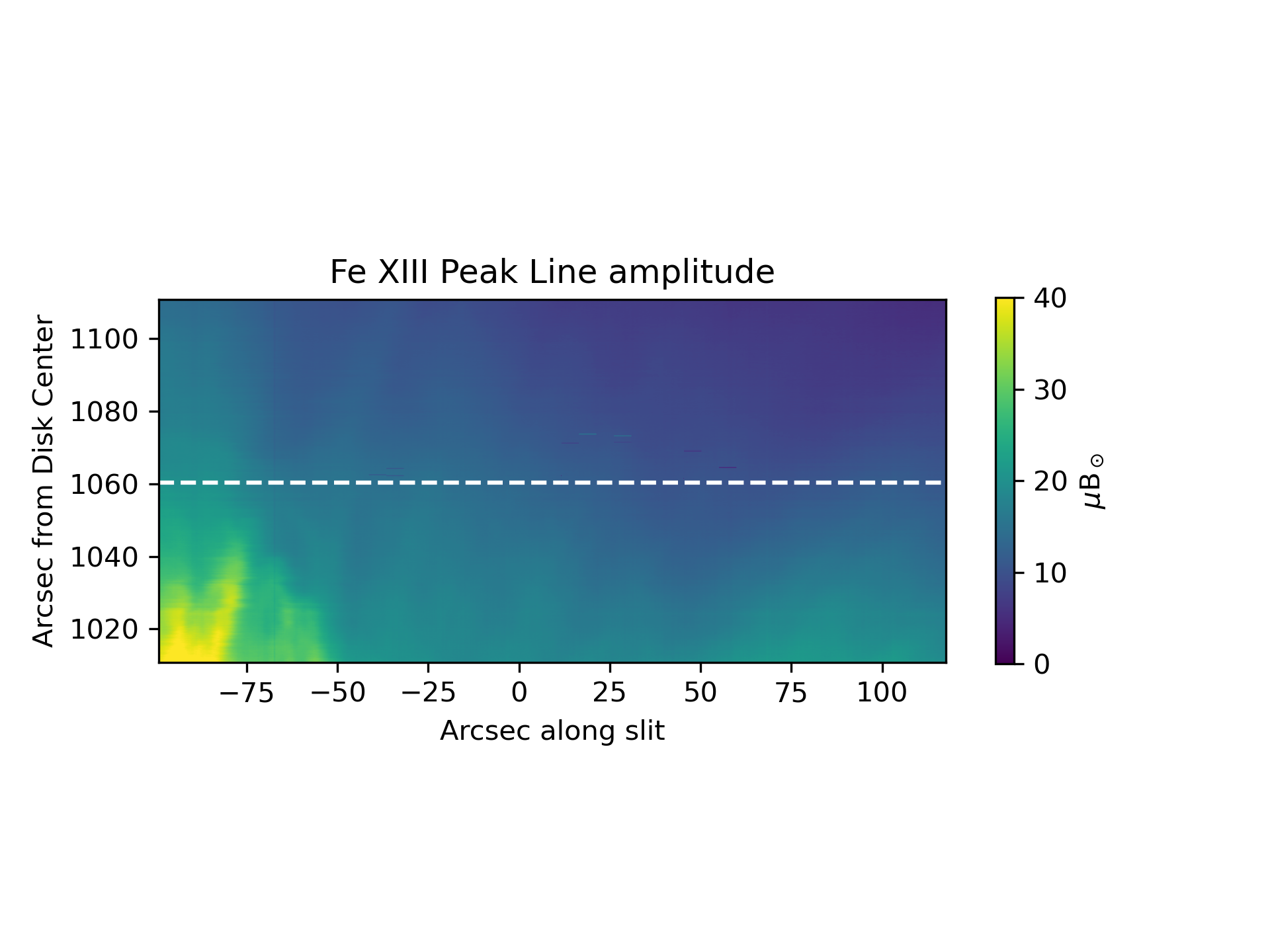}
    \caption{Results from Cryo-NIRSP raster scan. The upper panel displays the spectral profile (blue crosses) obtained by Cryo-NIRSP in the range (1072.5,1076)~nm and subsequent modelling. The full model fit to the observed spectrum is shown in green, while the Gaussian profile to the coronal emission line is shown in dashed purple. The bottom panel shows the 1074.7~nm line amplitude across the raster region highlighted in Figure~\ref{fig:overview}. The location of the sit-and-stare slit is indicated by the dashed line. Spectral radiance units are in millionths of the disk center intensity at the observed wavelength, $\mu B_\odot$.}\label{fig:raster-scan}
\end{figure}

\begin{figure}
    \centering
    \includegraphics[trim=250 0 0 0, clip, scale=0.6]{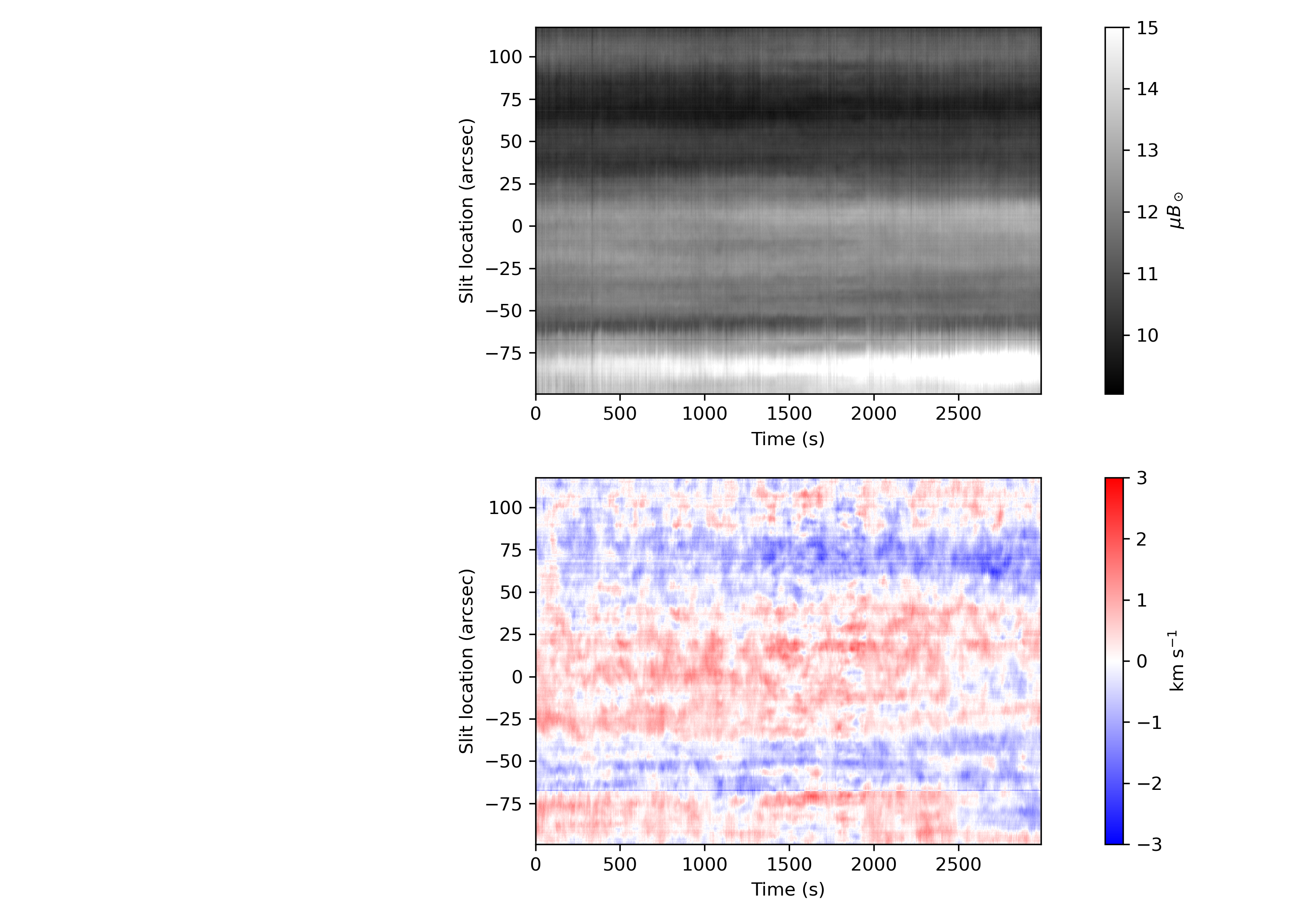} \\
    \caption{Sit-and-stare time-series. The top panel shows the 1074.7~nm line amplitude across the slit for each exposure in the sit-and-stare data. The bottom panel shows the corresponding Doppler velocity time-series. The mean value of Doppler velocity has been subtracted.}\label{fig:sit_and_stare}
\end{figure}

The observed target was the base of a coronal streamer, which, in the low corona, is typically comprised of a large system of loops. Previous CoMP observations have demonstrated these loop systems show some of the most clear signatures of propagating Alfv\'enic waves \citep{TOMetal2007,MORetal2016,sharma_2023}. The streamer is clearly visible in the SUVI data. We are fortunate that the Cryo-NIRSP slit actually straddles 3 distinct magnetic regions. The northern-most part of the slit ($<-50^{\arcsec}$ along the slit in Figures~\ref{fig:raster-scan} and \ref{fig:sit_and_stare}) overlaps the top of an active region NOAA AR13471, the middle section covers an open field region (seen mostly clearly in the filtered SDO/AIA 193~{\AA} image and its extension in the SUVI image) and the southern-most part ($>60^{\arcsec}$ along the slit in Figures~\ref{fig:raster-scan} and \ref{fig:sit_and_stare}) samples the streamer base. {We note that the position of DKIST field-of-view and Cryo-NIRSP slit is only approximate on the shown images.}

\medskip
For the Cryo-NIRSP data, the initial calibration steps were undertaken by the Cryo-NIRSP Science team, which largely consist of detector calibrations. Details of this process are discussed in \cite{Schad_2023,schad_2024}. An example spectra from the observations is shown in Figure~\ref{fig:raster-scan}. The coronal emission line is visible around 1074.7~nm, and there are a number of other spectral features relating to atmospheric absorption and instrumental scattering of disk radiation. In order to extract the coronal line profile we follow the method discussed in \cite{schad_2024}, where the background is modelled using the TCCON solar \citep{Toon} and HITRAN-modelled telluric \citep{schad_dkist}\footnote{Available at: \url{https://github.com/tschad/dkist_telluric_atlas}} spectral atlases and the coronal line profile is modelled with a Gaussian. Spectral profiles were fit for most pixels along the 246$^{\arcsec}$ projected slit length recorded on the detector; however, we eliminate $6^{\arcsec}$ at the upper edge due to calibration artifacts. The lower panel in Figure~\ref{fig:raster-scan} shows the line amplitude estimated from the raster-scan for the portion of the slit with valid spectral profiles. 

The data of primary importance for this study are shown in Figure~\ref{fig:sit_and_stare}, namely the time-series for the line amplitude and Doppler velocity (shown with mean value removed) from the sit-and-stare observation. It is seen in the line amplitude time-series that the emission from the open field and quiet Sun regions is relatively constant over the observation period. This indicates there is a thermal balance between heating and cooling of the plasma and no significant heating events occur, e.g., an energetic reconnection event. On the other hand, the active region emission is seen to increase over the period of observation. SDO/AIA images reveal that the active region is constantly varying, with loops appearing, growing and fading. This indicates substantial, episodic heating activity occurring in the region. 

\medskip
To support the analysis of the Cryo-NIRSP data, we also use a time-series of observations from SDO/AIA 193~{\AA} passband. The observations start at 20:00~UTC until 22:00~UTC. The time-series are obtained from the SDO Joint Science Operations Center cut-out service.

\section{Results}

\begin{figure*}
    \centering
    \includegraphics[trim=0 0 0 0, clip, scale=0.7]{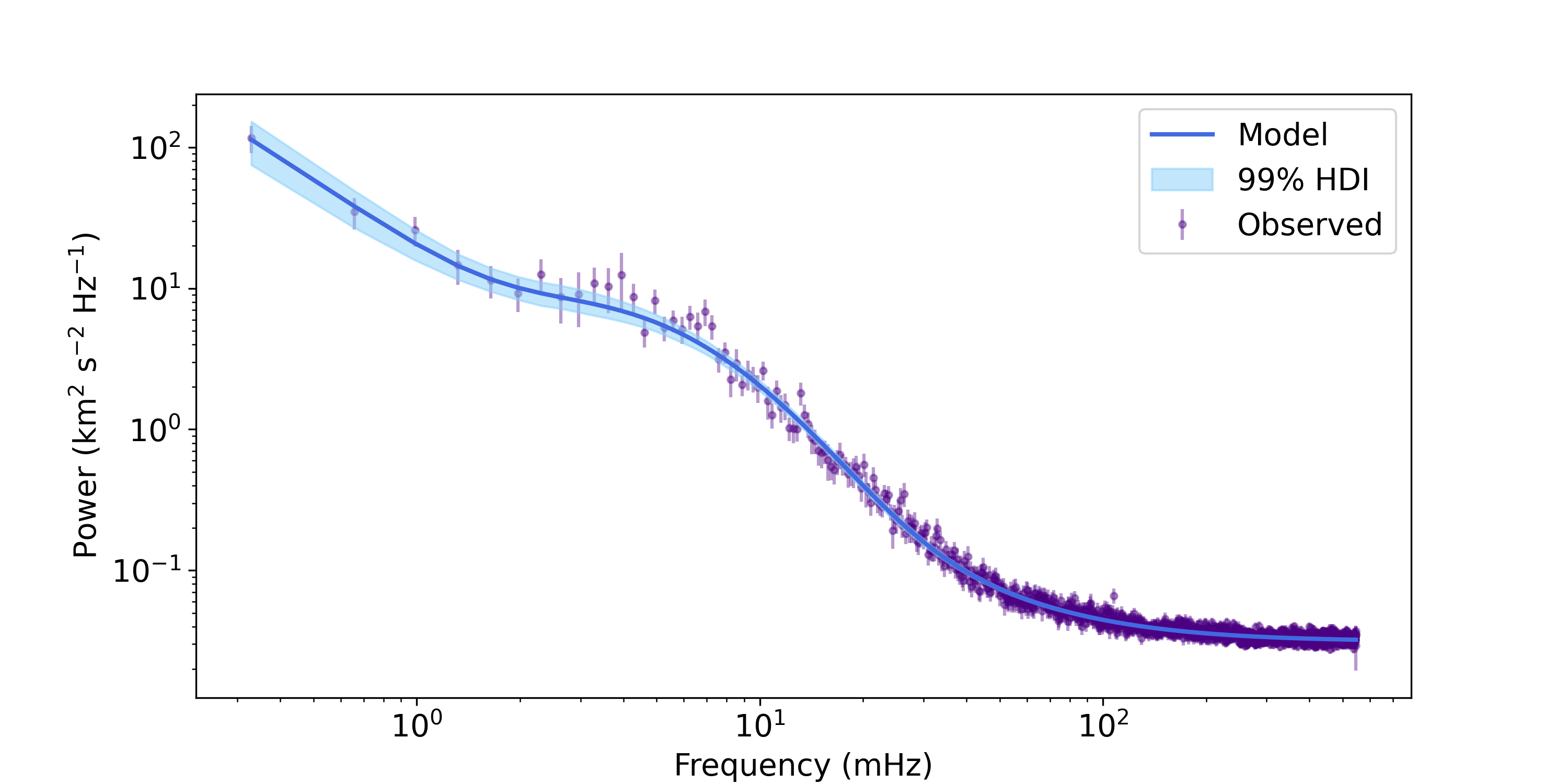}
    \caption{The power spectrum of coronal Doppler velocity fluctuations in the open field region. The estimated power spectrum is shown by the purple filled circles with corresponding estimated uncertainties. The dark blue line is the posterior mean estimate for the power spectrum model and the corresponding 99\% highest density interval (HDI) is shown as the filled light blue region.}\label{fig:power_spec}
\end{figure*}

\subsection{Power spectrum}
Our first step in the examination the Doppler velocity fluctuations is to estimate the power spectrum. Here we only show the results for the open field region crossed by the slit (between -50$^{\arcsec}$ and 60$^{\arcsec}$), but obtain similar results for the other regions. Each time-series is mean subtracted and multiplied by a Tukey (tapered cosine) window (with $\alpha=0.1$) before taking the discrete Fourier transform. In Figure~\ref{fig:power_spec} we show the average power spectrum from all the individual time-series in the open field region. It can be seen that it has similar features to those found in the CoMP observations, namely an excess of power around 4~mHz and power law behaviour. It can already be seen that the power law behaviour extends beyond 10~mHz, up until around 60~mHz where the power spectrum becomes noise dominated. To model the shape of the power spectrum and also examine the fluctuations beyond 60~mHz (see Section~\ref{sec:pow_spec_model}), further insights are needed about the Doppler velocity time-series. 

\subsection{Relationship between Doppler velocity and line amplitude}
Before focusing on the high frequency fluctuations, we want to establish whether the fluctuations seen in Doppler velocities have any relationship with the fluctuations in the line amplitude (i.e., comparing the two sets of time-series shown in Figure~\ref{fig:sit_and_stare}). Changes in line amplitude can occur due to variations in temperature and density of the plasma. From a wave perspective, any correlation between line amplitude and Doppler velocity signals would indicate that some of the Doppler velocity signals are associated with compressible MHD waves. Given the near-constant levels of emission over the observation period in the open field region and quiet Sun (Figure~\ref{fig:sit_and_stare}), we suggest any fluctuation in line-amplitude or Doppler velocity is probably due to wave processes occurring in a relatively stable plasma. 

\medskip

To examine the relationship between two time-series across a range of frequencies, one can use cross-spectral methods \citep[e.g.,][]{Lites_1993,Krijger_2001}. The Fourier transform of a time-series is given by
\begin{equation}
    F_1(f) = a_1(f) +ib_1(f), 
\end{equation}
where $a_1$ and $b_1$ are real numbers, and $f$ is the frequency. The cross-correlation spectrum between two time-series is
\begin{eqnarray}
     F_{12}(f) &=& F_1(f) \bar{F_2}(f),\label{eq:cc_spec} \\
               &=& c_{12}(f) + id_{12}(f), \nonumber 
 \end{eqnarray} 
where $c_{12}$ and $d_{12}$ are real numbers and $\bar{F_2}$ is the complex conjugate of the Fourier transform. The phase difference spectrum between the two time-series
is then given by
\begin{equation}
    \phi(f) = \arctan\left(\frac{d_{12}(f)}{c_{12}(f)}\right),
\end{equation}
and the magnitude squared coherence (MSC) is
\begin{equation}
    C^2(f)=\frac{\langle F_{12}(f)\rangle\langle \overline{F_{12}(}f)\rangle}{\langle F_{1}^2(f)\rangle\langle F_{2}^2(f)\rangle}.
    \label{eq:msc}
\end{equation}
The angular brackets denote that the MSC requires some form of spatial or temporal averaging, otherwise the coherence is unity regardless of any differences between the time-series. For pure noise signals, averaging over $n$ independent realisations will lead to an estimate of the coherence, $\mathbb{E}(C)=1/\sqrt{n}$. To calculate the phase and MSC we choose to use the Welch method \citep{welch1967}, which divides the time-series into overlapping segments and computes the cross-correlation spectrum, Eq.~\ref{eq:cc_spec}, for each segment and averages the spectra. Phase and MSC are then calculated with this average cross-correlation spectrum. While this reduces the frequency range and sampling, we believe it is preferred to a spatial averaging or Fourier smoothing. This is because the fluctuations have a spatial coherence over many pixels (see Section~\ref{sec:coher}), hence spatially local signals cannot be considered as independent realisations.

\begin{figure*}
    \centering
    \includegraphics[trim=0 0 0 0, clip, scale=0.7]{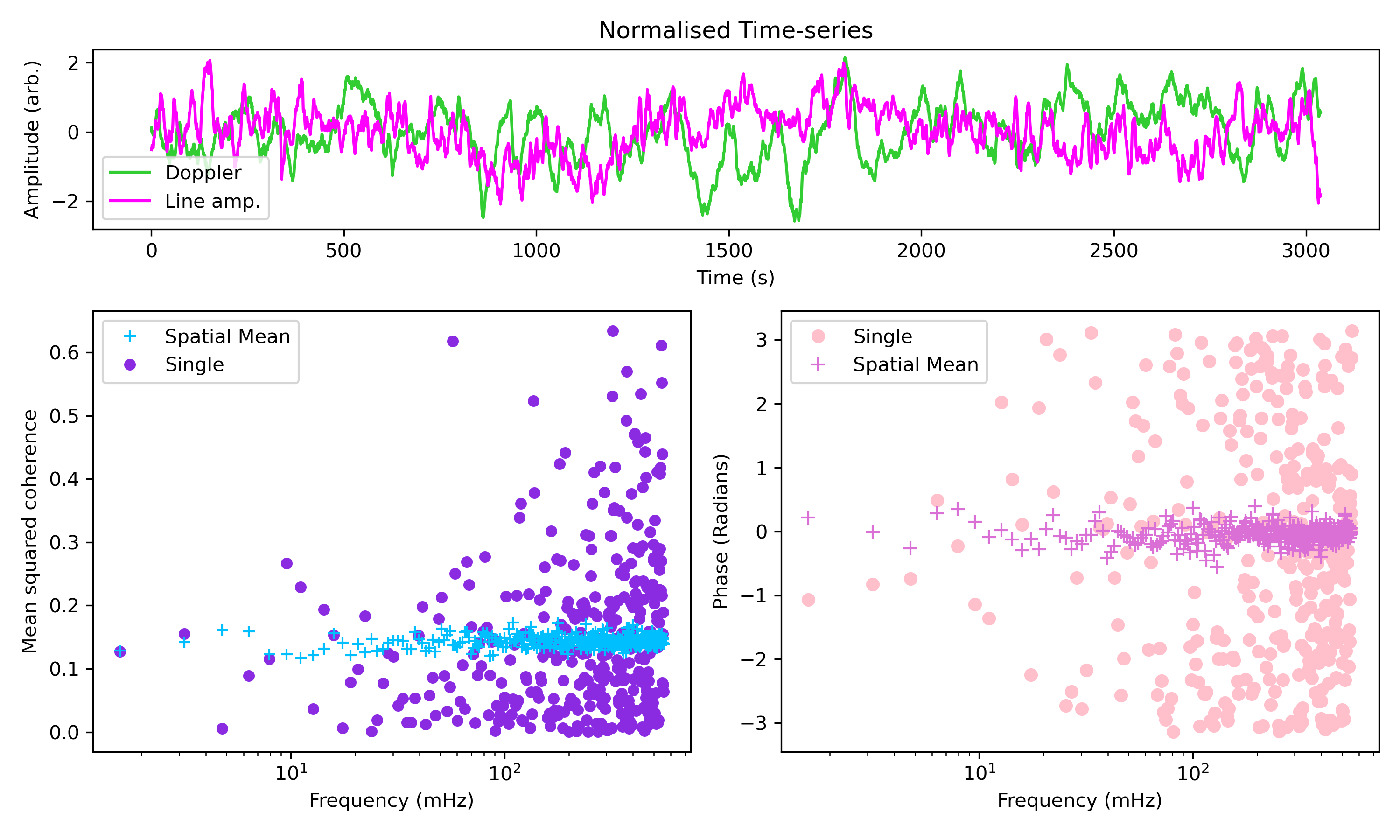}
    \caption{Comparison of line amplitude and Doppler velocity fluctuations. The top panel shows a pair of line amplitude and Doppler velocity signals from a single spatial location along the slit. The bottom left panel displays the magnitude squared coherence (purple dots) between the two times-series (estimated using Welch's method) and the spatial mean of all time-series pairs (blue crosses). The bottom right hand panel displays the phase for the same pair of time-series (pink circles) and the spatial mean (orchid crosses).}\label{fig:phase_coher}
\end{figure*}

This process is carried out for all valid pixels along the slit. In the top panel of Figure~\ref{fig:phase_coher} we show an example of a Doppler velocity and line amplitude time-series from the same spatial location. The two time-series have been normalised to aid comparison (for reference, typical line amplitude variations are $\delta I_{rms}\sim0.17~\mu B_\odot$ in locations outside the active region and $\delta I_{rms}\sim0.5~\mu B_\odot$ in the active region). Visual inspection of the time-series reveals that there is little correspondence between the fluctuations. This is confirmed by the results from the Fourier analysis. The lower panels of Figure~\ref{fig:phase_coher} show the MSC and phase results for an example single pair of time-series and the mean values from all pairs of time-series. It is seen that the individual estimates for both quantities are still relatively noisy, with large variations in values between neighbouring frequencies. Taking the spatial average demonstrates that the coherence between the signals across all frequencies is low, around 0.1, confirming the visual impression. The phase plot also shows there is a phase of zero when the results from all pairs of time-series are averaged. This occurs because, for each frequency, the phase values are randomly distributed between $(-\pi, \pi)$, suggesting there is no particular phase relation between the Doppler velocity and line amplitude signals. The lack of coherence and random phase demonstrate that the Doppler velocity and line amplitude fluctuations are essentially unrelated. Hence, we can be confident that the Doppler velocity fluctuations can be associated with incompressible dynamics, namely the Alfv\'enic waves.

\subsection{Spatial Coherence}\label{sec:coher}
The Doppler velocity signals observed in the corona are known to be spatially correlated. This is to be expected as they are thought to represent propagating Alfv\'enic waves, hence the wave signals should be coherent parallel to the magnetic field, i.e., the direction of wave propagation. It was also shown by \cite{sharma_2023} that the coronal Doppler velocity fluctuations were coherent perpendicular to the magnetic field, indicating that the fine-scale structure (typical size 100-2000~km) was not oscillating in isolation but with a collective motion over $\sim8$~Mm (this value is the mean of a wider distribution). \cite{sharma_2023} estimated this perpendicular correlation length for a narrow band of frequencies. Given the unique frequency resolution afforded by the Cryo-NIRSP data, we can estimate the correlation length-scale as a function of frequency. It is worth noting that, unlike \cite{sharma_2023}, we are unable to definitively determine the direction of wave propagation using these sit-and-stare data. A visual impression of the magnetic field direction can be obtained from processed AIA data (Figure~\ref{fig:overview}), and shows that the Cryo-NIRSP slit is likely only exactly perpendicular to a few of the over-dense plasma structures that outline the magnetic field, namely those in the open field region.  This means that the correlation length-scales are likely biased upwards from the perpendicular correlation lengths estimated by \cite{sharma_2023} (as correlation lengths along the magnetic field are longer).

To estimate the coherence lengths, we will again use the MSC given in Eq.~\ref{eq:msc}. We are interested in measuring the coherence for the full frequency range, so instead of using Welch's method we choose to average the cross-correlation spectrum in the frequency domain \citep[by using a box-car smoothing function of length 10 samples, e.g.,][]{Krijger_2001} before calculating the MSC. 

\begin{figure*}
    \centering
    \includegraphics[trim=0 0 0 0, clip, scale=0.55]{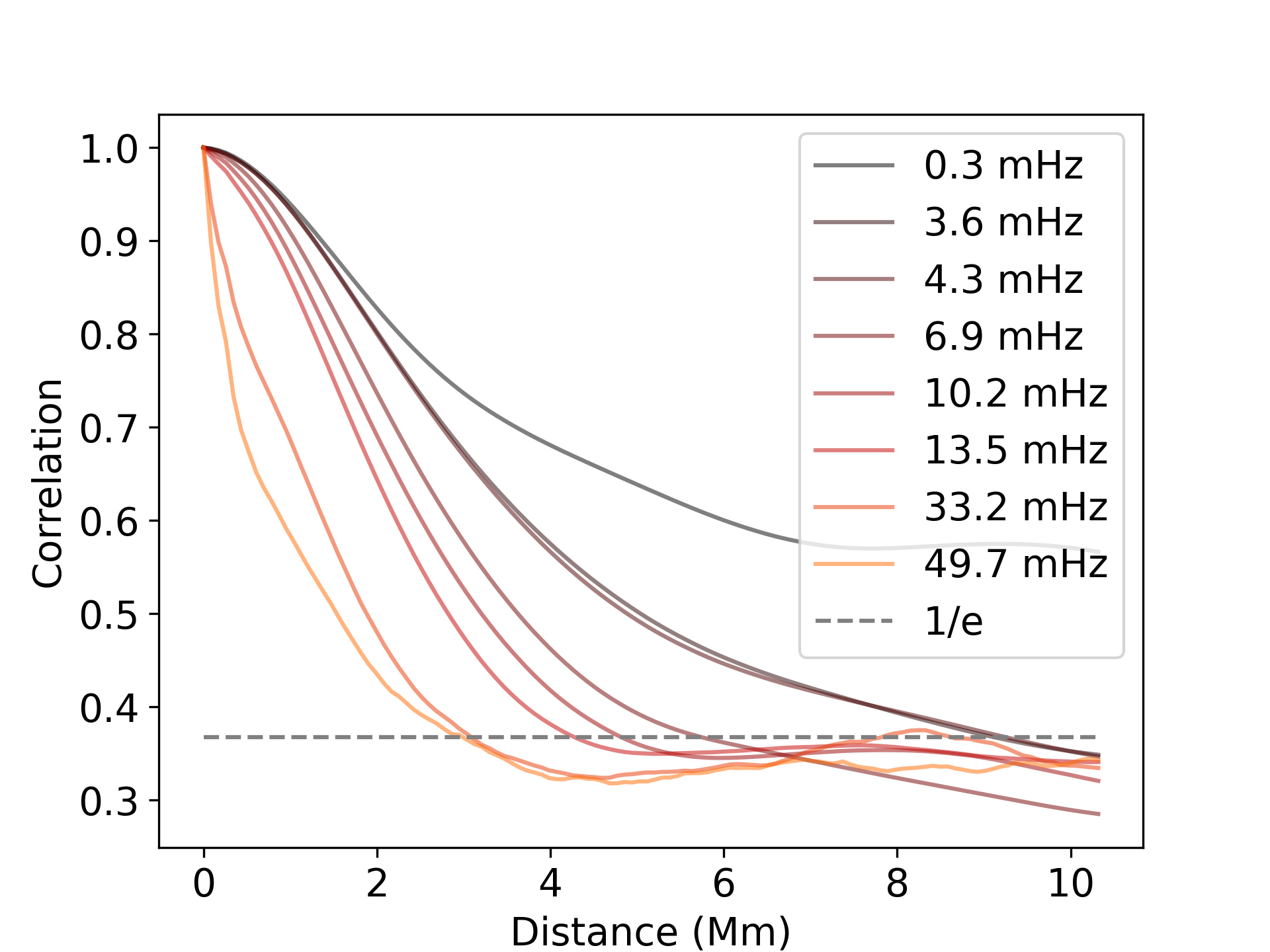}
    \includegraphics[trim=0 0 0 0, clip, scale=0.55]{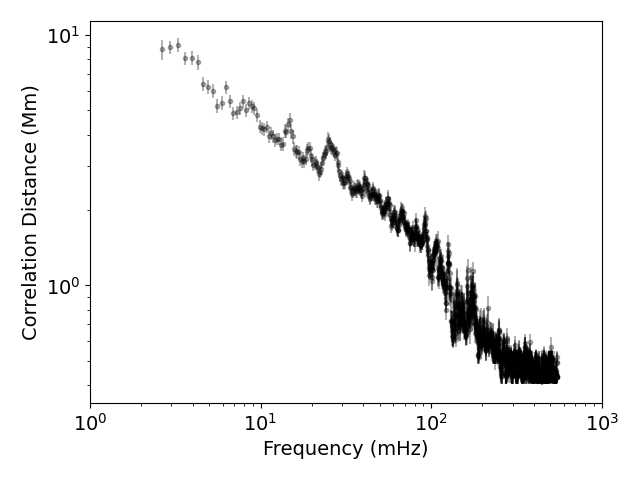}
    \caption{Spatial coherence of Doppler velocity time-series. The left panel shows the correlation curves as a function of distance for different frequencies. The curves are mean values from all pairs of time-series. The horizontal dashed line indicates the $1/e$ value. The right hand panel shows the spatial distance at which the correlation curves reach the $1/e$ value for each frequency. Each point is the median value of the samples across the slit and the uncertainties are the standard deviation of values calculated from bootstrapping.}\label{fig:spatial_msc}
\end{figure*}

Pairs of time-series separated by a distance $\delta x$ along the slit are compared, calculating the MSC as a function of frequency for all pairs. We then estimate the correlation length as the distance $\delta x$ for which the square root of the MSC drops below $e^{-1}$. In the left panel of Figure~\ref{fig:spatial_msc} we demonstrate how the coherence (or correlation) varies as a function of distance for different frequencies. The right panel of Figure~\ref{fig:spatial_msc} shows the estimated correlation lengths for all frequencies. It is seen that for frequencies around 4~mHz the coherence extends to around 6-8~Mm, in-line with the previous results from CoMP data \citep{sharma_2023}. The results here show a more nuanced picture and the correlation length decreases with increasing frequency (Figure~\ref{fig:spatial_msc} right panel). From a statistical point of view, this means that fluctuations at higher frequencies are more independent from signals in neighbouring pixels than those with lower frequency waves. When it comes to the averaging of power spectra across the slit, then the samples of power estimates from the DFT at higher frequencies will contain more independent samples than the estimates for power at lower frequencies. As a note, for frequencies smaller than $1.6$~mHz, the correlation length analysis is less reliable as there are fewer unique frequency values within the box-car average window. Furthermore, it can be seen that the coherence length-scale levels out around 0.2~Hz, reaching a value of $\sim$440~km (0.6$^{\arcsec}$). We take this value as an indication of the seeing-limited resolution of the data, where neighbouring pixels are coherent due to the local convolution of a point spread function (i.e., image blur). 

\subsection{Model of power spectrum}\label{sec:pow_spec_model}
In order to uncover the behaviour of the highest frequency fluctuations in the data, we need to estimate and remove the power arising from the data noise in the power spectrum. The data noise is assumed uncorrelated and white, which implies a constant level of power across all frequencies. From the estimated power spectrum shown in Figure~\ref{fig:power_spec}, it is seen that the white noise level starts to dominate the spectra at around 60~mHz and beyond. The white noise power is also additive hence we can treat the estimated power spectrum as a simple sum of the true spectrum and the noise spectrum. Once the level of white noise is known, it can be subtracted from the estimated power spectrum. To estimate the noise level, and characterise the distribution of power, we will utilise a Bayesian approach to model the power spectrum.

\medskip

The Bayesian framework used to model the data is summarised by: i) the priors
\begin{eqnarray}
    A&&\sim \mbox{HalfNormal}(1),\nonumber\\
    B&&\sim \mathcal{N}(1,0.5),\nonumber\\
    C&&\sim \mathcal{U}(2,6), \nonumber\\
    D&&\sim \mathcal{N}(10,2), \nonumber\\
    \mu &&\sim \mathcal{N}(-5.5,0.1),\nonumber\\
    \sigma&&\sim \mathcal{U}(0,0.8),\nonumber
\end{eqnarray}
where $\mathcal{N(\mu,\sigma)}$ is the Normal distribution and
$\mathcal{U}(a,b)$ is the continuous Uniform distribution defined on the interval $[a,b]$; ii) the model
\begin{eqnarray}
    S(f_i) &&= Af_i^{-B}+\exp(C)+\nonumber\\
        &&\exp(D)\exp\left[-(\log(f_i)-\mu)^2/(2\sigma^2)\right];\nonumber
\end{eqnarray}
iii) the likelihood function
\begin{eqnarray}
    P(S|f_i,A,B,C,D, \mu,\sigma) &&\sim \mathcal{N}(S(f_i),\epsilon_i).\nonumber
\end{eqnarray}
Here our model for the true power at frequency $f_i$ is given by $S(f_i)$, which is comprised of a power law component and a log-normal function. Such a model has been used previously to fit the power spectrum of Doppler velocity time-series in CoMP data \citep{MORetal2019}. The likelihood function is characterised by a Normal distribution with uncertainties $\epsilon_i$ for each estimate of power $\hat{S}(f_i)$ (details on the calculation of uncertainties are given Appendix~\ref{app:uncert}). The Bayesian model parametrises the white noise level as $\exp(C)$, and the amplitude enhancement as $\exp(D)$. This is implemented so that $C$ and $D$ are a similar order of magnitude to the rest of the parameters, which helps the Markov Chain Monte Carlo sampler. 

The prior distributions are not uninformative and were chosen to exclude certain unphysical cases. The chosen priors are still only weakly informative, providing little information to the posterior. This is confirmed by prior predictive checks which demonstrates the prior model only weakly identifies the observational model \citep{Gabryetal2019}. To sample the posterior we use a Hamiltonian Monte Carlo (HMC) sampler \citep[making use of PyMC -][]{pymc}, running multiple independent sampling chains.

In Figure~\ref{fig:power_spec} the result of fitting the above model to the data is shown. The most probable model is shown by the blue curve and the light blue shaded region shows the 99\% highest density interval (HDI) for the model. It can be seen that the model is well constrained by the data, with a very narrow HDI interval, especially at higher frequencies. The individual parameter estimates for the model are given in Table~\ref{tab:par_est}. We note that the model was fit to the square of the absolute values of Fourier coefficients, which is what the values given for the prior distributions correspond to. The given parameters in Table~\ref{tab:par_est} are reported for the power spectral density, which is different by a factor of $2/f_s/s$, where $f_s$ is the sampling frequency and $s=\sum w_i^2$ with $w_i$ the values of the window function. 

{In general the model describes the data reasonably well. Although, there appears to be some discrepancy between the model and apparent relatively narrow peaks around 4~mHz and 7~mHz. The power law component of the model is discussed in detail in Section~\ref{sec:high_freq_sub}. The power enhancement captured by log-normal function is relatively broad in frequency space, sitting above the power law between 1-50~mHz (see also Figure~\ref{fig:high_freq}). The extent of the enhancement is bigger than perhaps expected from earlier studies of the coronal power spectra \citep[e.g.,][]{MORetal2019}, although the scale factor ($\sigma$) in those studies was poorly constrained by the CoMP data. The peak of the log-normal function is $3.5\pm{0.5}$~mHz consistent with the earlier studies. While we will not focus on this topic further here, it is worth mentioning that the breadth of the power enhancement is much broader than that expected from the mode conversion of \textit{p}-modes \citep[e.g.,][]{Miriyala_2025}. However, the additional narrow peak at 4~mHz is more inline with expectations for the mode conversion process. The peak at 7~mHz is unexpected but coincides with significant time-scales recently identified in the Parker Solar Probe data \citep{Huang__2024}.}

Figure~\ref{fig:high_freq} shows the spectrum with the constant noise level (indicated by $exp(C)$ in the model) subtracted out. For additional discussion of this high-frequency part of the spectrum, see Section~\ref{sec:high_freq_sub}.

\begin{table}[ht]
\begin{tabular}{l|l|l|l}
    & Mean & 3\% HDI & 97\% HDI \\
 \hline
 $A$& 3.5$\times10^{-4}$& 3.0$\times10^{-4}$ & 4.0$ \times10^{-4}$ \\ 
 $B$& 1.58 & 1.53 & 1.63   \\
 $\exp(C)$& 0.0313 & 0.0311 & 0.0316   \\
 $\exp(D)$&  4.8&  3.8 & 5.8  \\
 $\mu$ & -5.66 & -5.80 & -5.52   \\
 $\sigma$& 0.693 & 0.651 & 0.736   
\end{tabular}
\caption{Marginalised posterior parameter estimates for model fit to power spectrum. The uncertainties are taken from the 94\% highest density interval (HDI).}
\label{tab:par_est}
\end{table}

\section{Discussion}

\subsection{Incompressibility}\label{sec:sub_incomp}
In the present observations, the coronal Doppler velocity fluctuations are found to represent incompressible dynamics, showing no relation to the corresponding line amplitude fluctuations. The lack of any correlation of Doppler velocity with line amplitude might strike some as counter-intuitive, however, these results are in line with properties of slow magneto-acoustic modes observed previously in the corona. The presence of compressive fluctuations has typically been inferred through variation in the intensity of coronal emission \citep[e.g.,][]{DEFGRU1998,De_Moortel_2002}. The slow waves are found to be upwardly propagating (suggesting an origin in the lower atmosphere) and have been found to damp rapidly with height, with typical damping lengths of $\lesssim20$~Mm \citep{De_Moortel_2002,Marsh_2011,Meadowcroft_2023}. Although results from \cite{Mandal_2018} suggest slow waves in coronal plumes may have a greater damping length, on the order of $40$~Mm. So it should be anticipated the compressive modes are significantly damped by 0.1~$R_\odot$ ($\sim 70$~Mm).

The Alfv\'{e}nic waves themselves may also excite compressible waves in-situ through shocks and the parametric decay instability \citep[e.g.,][]{Suzuki_2006}. However, this requires non-linear Alfv\'{e}nic waves, which do not appear to be present at the heights observed here. Typical waves amplitudes are around 15-30~km~s$^{-1}$ \citep{MCIetal2011,THUetal2014,MORetal2019} while Alfv\'en speeds are on the order of 200-1000~km~s$^{-1}$ in the quiet Sun and coronal holes \citep{MORetal2016,YANGetal2020}. Hence normalised wave amplitudes are on the order of 0.01-0.15 and the Alfv\'enic waves at $\sim1.1$~$R_\odot$ can be considered nearly linear. This agrees with the results from wave-driven compressible solar wind simulations, which predict little to no compressible fluctuations at these heights \citep[e.g.,][]{Suzuki_2006,Shoda_2019,Matsumoto_2020}.

\begin{figure}[!t]
\centering
    \includegraphics[trim=0 0 0 0, clip, scale=0.7]{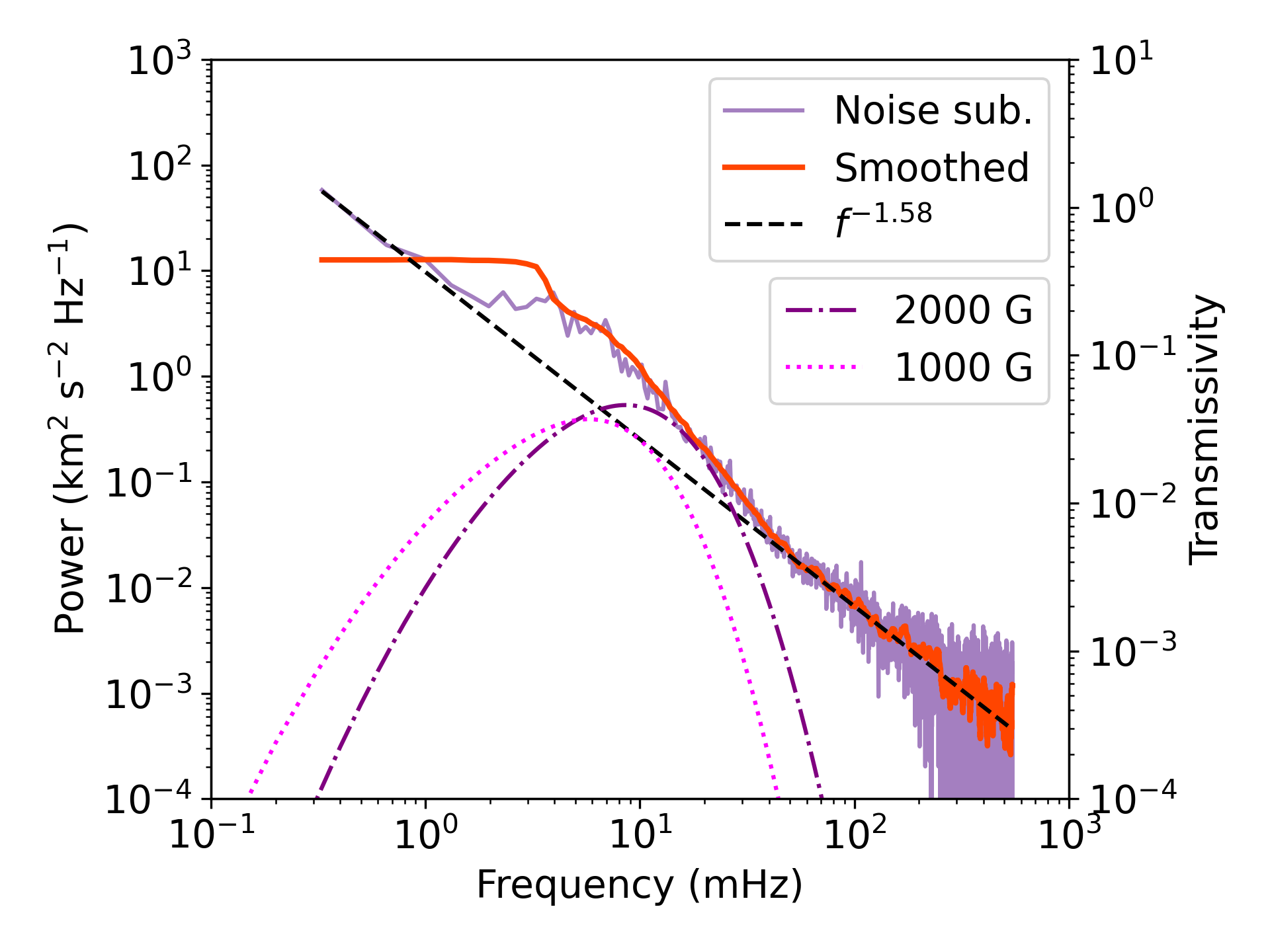}
    \caption{Extended power law to high frequencies. Displayed in the figure are the power spectrum from Figure~\ref{fig:power_spec} with the constant noise level removed (indigo curve) and the $f^{-1.58}$ slope (posterior mean estimate for slope, black dashed line). A smoothed version of the same power spectrum is overplotted for visualisation (orange). Also shown are two suggested curves for the coronal transmission profile of Alfve\'n waves suggested by \cite{2019ApJ...871....3S}, calculated for photospheric magnetic field strengths of 2000~G (purple) and 1000~G (magenta).  }\label{fig:high_freq}
\end{figure}

\medskip

\begin{figure}
    \centering
    \includegraphics[trim=0 50 0 60, clip, scale=0.8]{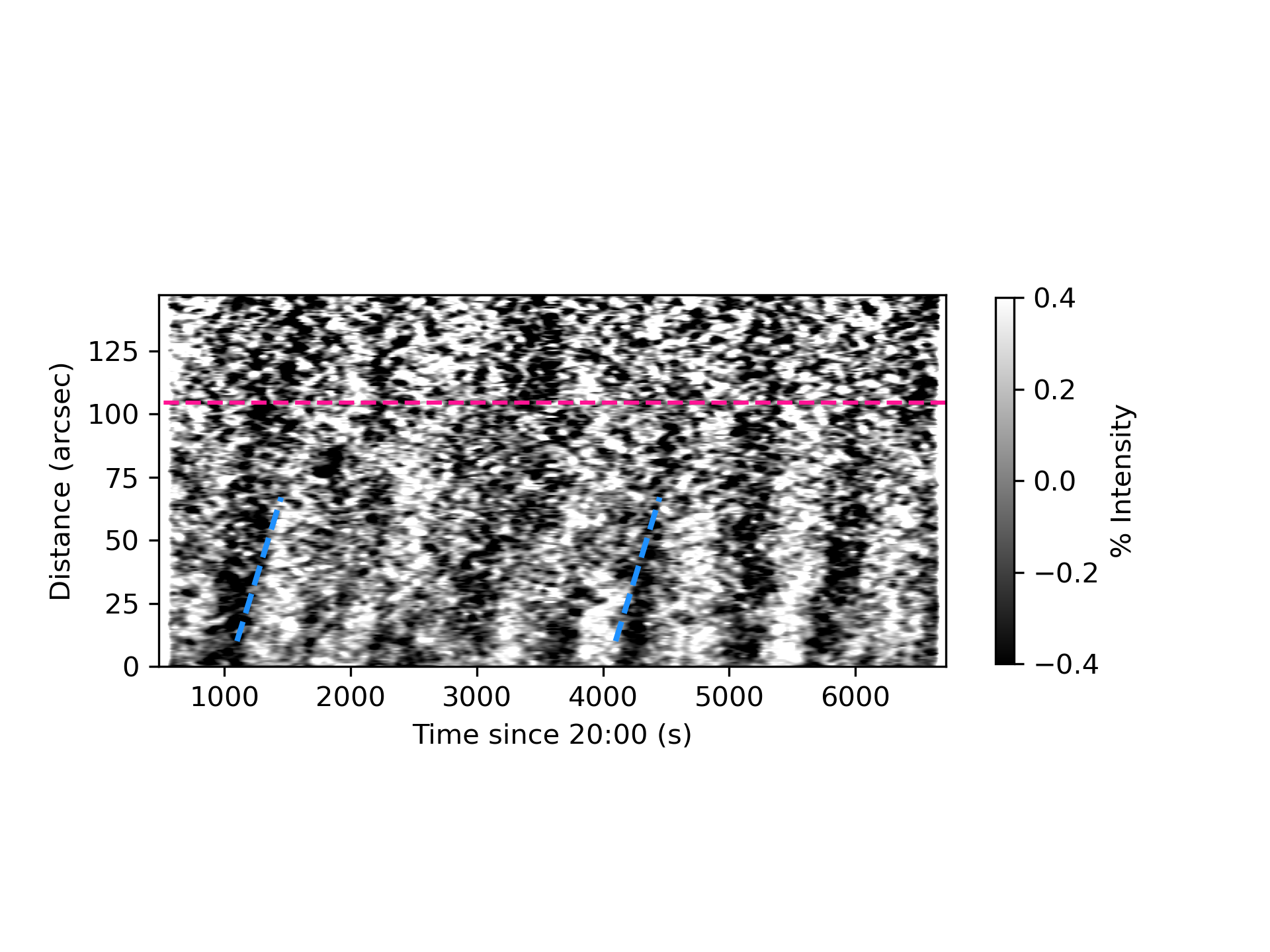}
    \caption{Time-distance diagram generated from SDO/AIA 193~{\AA} observations. The vertical axis is distance along the line indicated by "TD slit" in Figure~\ref{fig:overview}, which follows the fine-scale structure in the open field region. The approximate location of the Cryo-NIRSP slit is marked by the pink dashed line. The two blue, dashed lines indicate propagation speeds of 120~km/s. }\label{fig:aia_comp}
\end{figure}

To provide an additional perspective on the presence of compressible waves, we have analysed a time-series of SDO/AIA images from the 193~{\AA} passband.  The emission from this passband depends primarily on the square of
the electron density at coronal temperatures ($\log T \sim 6.0$)
and is completely independent of Doppler velocity. The data were processed with Fourier filtering to remove noise \citep{DeForest_2017} and established methods to identify compressible motions were applied to the data (e.g., running difference - \citealp{Mandal_2018}; or time-normalisation - \citealp{Morgan_2018}). We find evidence for compressive fluctuations, although only for small-amplitude, long-period waves. An example of the analysis is shown in Figure~\ref{fig:aia_comp}, which presents a time-distance diagram that follows the path of the fine-scale structure seen in filtered images. The time-distance diagram is centred on the location indicated in Figure~\ref{fig:overview} (coinciding with a location near the centre of the Cryo-NIRSP slit shown in Figure~\ref{fig:sit_and_stare}) but is averaged over a region of 13$^{\arcsec}$ perpendicular to the shown line. For this example, the SDO/AIA data is subject to a running-difference filter with a sliding window of 480~s. The choice of window size made little difference to the results. Figure~\ref{fig:aia_comp} shows evidence for upwardly propagating periodic features that are often reported in fan loops or open field regions. The periodicity is $\sim20$~minutes, the signals propagate at $\sim120$~km/s with an amplitude less than 1$\%$ of the background intensity. It can also be seen to decay with height, with the signal apparently disappearing by the time it reaches the height at which we observe with Cryo-NIRSP. Taken together with the results from the cross-spectral analysis of the Cryo-NIRSP data, the results from SDO/AIA indicate compressible fluctuations may be present but do not contribute significantly to the Doppler velocity fluctuations. We caution that this result is not necessarily general and significant correlations between Doppler and line amplitude fluctuations might be present elsewhere in the corona. As noted, at lower altitudes upwardly propagating compressible fluctuations have been found previously and at higher altitudes non-linear Alfv\'enic wave begin to generate compressible fluctuations.

\subsection{Source of high-frequency fluctuations?}\label{sec:high_freq_sub}
Of particular interest here is the nature of the high-frequency component of the fluctuations, namely above 10~mHz. The low-frequency properties of the fluctuations have been defined previously \citep[e.g.,][]{MORetal2016} and the Cryo-NIRSP data demonstrates the same behaviour, i.e. a power law at low frequencies and an enhancement of Doppler velocity power around 4~mHz. After removing the white noise component from the estimated power spectrum, Figure~\ref{fig:high_freq} demonstrates that the Doppler velocity fluctuations continue to follow a power law down to the highest frequencies ($\sim1$~Hz). The estimated power law is $f^{-1.58}$, and the index falls within the range of values reported in previous analysis of CoMP data \citep{MORetal2019}. Despite this seemingly innocuous result, the continuation of the power law to higher frequencies challenges various models of Alfv\'enic wave generation and propagation.

As noted in the introduction, numerous works have shown that the damping of Alfv\'enic waves excited in the photosphere is stronger at higher frequencies due to ion-neutral effects, resonances and/or phase mixing in the chromosphere \citep[e.g.,][]{2011A&A...534A..93Z,Soler_2015b,2017ApJ...840...20S,2019ApJ...871....3S}. In Figure~\ref{fig:high_freq} we show the estimated transmission profile for Alfv\'en waves from \cite{2019ApJ...871....3S} (see Appendix~\ref{app:trans} for an extended discussion on the choice of profile). The transmission profile accounts for various wave damping mechanisms (e.g., phase mixing, ambipolar diffusion) and reflection in the lower solar atmosphere. The profile depends upon the magnetic field strength and we show two curves calculated using typical values of photospheric magnetic fields associated with quiet Sun network elements \citep[i.e., from measurements of magnetic bright points;][]{BERTIT2001,ISHetal2007}. The transmission profile shows a significant drop in wave transmission below 10~mHz which is due largely due to ion-neutral effects and efficient phase mixing causing significant damping. Clearly the observations here suggest substantially more power in the high frequency incompressible fluctuations than is predicted to arise from upwardly propagating Alfv\'enic waves excited by photospheric motions.

\medskip

Assuming that the predictions of strong chromospheric damping of Alfv\'enic waves excited at the photosphere are realistic, the extension of the coronal power law to high-frequencies then naturally raises the question: what is the source of the high-frequency fluctuations?

\medskip

One potential option is magnetic reconnection, which has been invoked in the past as a way to generate both high-frequency waves \citep[with frequencies greater than 1~Hz;][]{axford1992,MarchTu1997} and low-frequency waves \citep[with frequencies less than $10^{-3}$~Hz;][]{Cranmer_2018}. Magnetic reconnection and flux cancellation is now readily reported across a range of scales, with frequent, relatively small-scale events leading to various jet-like phenomenon in the corona \citep{Raouafi_2014,Kumar_2022,wang_2022,Raouafi_2023, Chitta_2023}. The jets occur on scales of 100-2000~km with typical time-scales 20-300~seconds, which might also be indicative of the scales of any associated wave motion. The jet-like events are suggested to be significant contributors of mass to the solar wind and are linked with upwardly propagating disturbances seen in polar regions \citep{Raouafi_2023}. However we do not notice any of these events in the SDO/AIA data for this region during the observations, although some of the jet activity may be occurring on scales smaller than AIA can resolve \citep{Chitta_2023}. Moreover, \cite{Raouafi_2023} estimated that there are $10^5$ small-scale jets per day across the Sun, which is $\sim0.003$~hr$^{-1}$~Mm$^{-2}$. Jets on smaller scales, the so-called picoflare jets, appear to be more prevalent but still only suggest activity at rate of $\sim0.03$~hr$^{-1}$~Mm$^{-2}$ \citep[based on jets reported in][]{Chitta_2023}. This rate of jet production would seemingly not sustain the wave activity seen.

Another source of the Alfv\'enic waves is chromospheric jets or spicules. Simulations of spicule formation by \cite{MARetal2017} show that Alfv\'enic waves are generated along with the spicule due to the release of magnetic tension. Typical periods of the waves in the simulation are 30-150~s, which would only contribute to a portion of the power spectrum. These values are consistent with observations of Alfv\'enic waves in chromospheric spicules and fibrils \citep[e.g.,][]{JESetal2015, Bate_2022}. There is also observational evidence that magnetic reconnection can drive spicules \citep[e.g.,][]{2019Sci...366..890S} which would also generate waves as a by-product \citep[e.g.,][]{Takeuchi_2001,Thurgood_2017}. Spicules are also ever present on the sun, with estimates ranging from $10^5$ \citep{Sekse_2012} to $10^7$ \citep{JUDCAR2010} spicules at any one time. Their number density then ranges from 0.02-2~Mm$^{-2}$ and they occur preferentially along the boundaries of the network, hence are present at the base of many coronal structures. {The presence of the propagating disturbances found in the SDO/AIA is thought to be an indication of slow modes excited by chromospheric jets \citep[e.g.,][]{Samanta_2015}. The propagating disturbances appear to be relatively infrequent (around one every 15-20 minutes), although this may not be an indication of how often chromospheric jets produce Alfv\'enic waves.}

\medskip

A final physical option must also be mentioned and that is turbulence. MHD turbulence can cascade wave energy from low wavenumbers to high wavenumbers. It is known to be anisotropic in nature, suppressing the cascade in the direction parallel to the magnetic field \citep{2003LNP...614...28O,Schekochihin_2022}. The critical balance theory \citep{Goldreich_1995} suggests strong MHD turbulence exhibits sufficient mixing between the turbulent motions (mostly perpendicular to the field) and the flow of Alfv\'{e}n wave packets (parallel to the field) such that their timescales remain coupled together at different spatial scales:
\begin{equation}
  \omega \, = \, k_{\parallel} V_{\rm A}
  \, \approx \, k_{\perp} v_{\perp}
  \label{eq:critbal}
\end{equation}
where the frequency, $\omega$, and parallel wavenumber, $k_{\parallel}$,
refer primarily to a conception of the fluctuations as waves,
and the scale-dependent perpendicular velocity, $v_{\perp}$, and
perpendicular wavenumber, $k_{\perp}$, refer mainly to a conception of the fluctuations as short-lived turbulent eddies. Critical balance suggests that the perpendicular cascade follows a power law of $k_\perp^{-5/3}$ while the parallel cascade is $k_\|^{-2}$. 

Interpreting the observed frequency power spectrum is not as obvious as it may seem. The nature of our observations, in principle, provides an examination of the system from a Eulerian perspective. We observe at a point and measure the fluctuations in the flow that is passing by. \cite{Beresnyak_2015} suggests that the corresponding parallel spectrum can be estimated from the Lagrangian frequency spectrum, as the Eulerian one can be dominated by sweeping effects (where the larger eddies transport the smaller ones). The picture for turbulent systems, however, appears more subtle than this. \cite{Lugones_2016,Lugones_2019} demonstrate that for systems with strong mean magnetic fields and/or large cross-helicity, then the dynamics are dominated by the Alfv\'enic waves for a broad range of scales (rather than the sweeping). For Alfv\'enic waves, the relationship between velocity and magnetic perturbations is $b_\perp=-B_0v_\perp/c_{ph}$ (where $v_\perp$ is the amplitude of the velocity fluctuations perpendicular to the field, $c_{ph}$ is the phase speed), hence in the corona, $|b_\perp|\lesssim 0.1B_0$ with $v_\perp=15-30$~km~s$^{-1}$ and $c_{ph}=400$~km~s$^{-1}$ \citep[suitable for an open field region][]{MORetal2015}. Hence, we expect to be in strong field regime in the low corona for the frequencies typically observed. This is supported by previous analysis of CoMP observations that have shown the Alfv\'enic fluctuations follow a dispersion relation satisfying $\omega=c_{ph}k$ \citep[at least up to frequencies of 10~mHz;][]{TOMMCI2009,MORetal2015,tiwari_2021}. Given these insights from previous observations, it would seem natural to interpret the observed frequency power spectrum as one that can be converted to the parallel spectrum, hence would suggest $k_\|^{-1.58}$. Clearly this is different from the expected exponent for the parallel spectrum from critical balance theory (and is actually closer to that expected for the perpendicular one, see Appendix~\ref{app:scales} for further discussion). 

This does not preclude turbulence though, as there are other
proposed scaling laws besides critical balance (see Appendix~\ref{app:scales})
or the turbulence observed in the corona could be in a state of
development. The fluctuations in the lower solar atmosphere can potentially be in a fully turbulent state, with the reflection of Alfv\'enic waves generated in the photosphere occurring due to the increase in Alfv\'en speed with height in the lower atmosphere. Significant levels of turbulence are found in the chromosphere in various numerical models of wave turbulence that include the lower solar atmosphere  \citep[e.g.,][]{VANBALLetal2011, Matsumoto_2012}. The fluctuation spectrum predicted from phenomenological turbulence theories will likely be modified by frequency-dependent wave damping that arises from perpendicular gradients in the Alfv\'en speed (e.g., phase mixing, resonance), as well as partial ionisation. Moreover, any spectrum set in the lower solar atmosphere will also be modified on transmission to the corona, altered by the frequency-dependent reflection from the transition region. Hence, the local (variation in $v_A$ perpendicular to the field) and large-scale inhomogeneities (significant variation in $v_A$ along the field) then disrupt the fluctuations from the fully developed state, and it will take time for the fluctuations to recover. There is evidence that the corona reaches a fully developed state of turbulence by at least 3~$R_\odot$ \citep{Telloni_2024}.

Alternatively, a different model of turbulence aside from critical balance might be more appropriate description for the wave dynamics. For example, the nearly incompressible MHD turbulence predicts a $k_\|^{-5/3}$ in system with large cross-helicity \citep[i.e., predominatly outwardly propagating fluctuations;][]{Zank_2020}. This description of turbulence has support from the analysis of particular solar wind streams, were a $k_\|^{-5/3}$ index is estimated from the magnetic power spectrum \citep{Telloni_2019}. 

\medskip
A final alternative to sources of solar origin is one of terrestrial origin, which cannot be completely ruled out. There is a chance that the signal is correlated noise arising from the terrestrial atmosphere or the instrumentation. As of yet we cannot identify a specific source. It could be that we have to wait until the launch of a space-based instrument with high cadence spectroscopic capabilities (e.g., MUSE or Solar-C) to rule out this option. 

\subsection{Coherence}\label{sec:coher_discuss}
In the main text we use the coherence of the fluctuations along the slit to aid our uncertainty estimates for the power spectrum. However, the measure of coherence is interesting in its own right. From the fine-scale structure revealed in the SDO/AIA images (Figure~\ref{fig:overview}), it can be seen that the slit is nearly perpendicular to the magnetic field direction in the open field region. Hence, the coherence values can be associated with the perpendicular scales of the fluctuations. 

Previously, \cite{sharma_2023} investigated the fluctuation scales perpendicular to the magnetic field. They used CoMP data, which had the benefit of two spatial domains and the ability to ascertain the field direction. Their analysis was restricted to the waves with frequencies centered around $4$~mHz and found scales of $\sim8$~Mm. This is consistent with our results for the lower frequencies (2-5~mHz). {We also note these values are similar to estimates from \cite{Bailey_2025}.} 

Interestingly, Cyro-NIRSP enables a more nuanced study of the perpendicular scales. Here, it is found that the correlation scales of the fluctuations decreases with increasing frequency (Figure~\ref{fig:spatial_msc}). This behaviour is predicted from MHD turbulence theory. Depending upon the particular description of strong turbulence, it can be shown that there is a predicted relationship between the parallel and perpendicular scales, namely $k_\|\propto k_\perp^{1-n}$ (where $n$ is a scaling arising from the theoretical description). Hence, assuming the correlation distance behaves essentially as the inverse wavenumber, then it can be expected that $L_{corr}\propto\omega^{1/(n-1)}$ (see Appendix~\ref{app:scales}). For scale-dependent dynamic alignment \citep{boldyrev_2006}, with $n=1/4$, then $L_{corr}\propto\omega^{-1.33}$. This is significantly steeper than the observed slope in the right panel of Figure~\ref{fig:spatial_msc}. Other predicted scaling exponents from strong turbulence descriptions predict even steeper slopes.

There is some caution required with measuring the scaling between correlation distance and frequency in Figure~\ref{fig:spatial_msc}. As noted in Section~\ref{sec:coher}, for the Cryo-NIRSP exposure times, the effect of atmospheric turbulence will be to convolve the coronal emission lines with some local PSF across the slit. This leads to signals in neighbouring pixels to be correlated. The results from the correlation distance plot suggest that there is a lower limit of 0.6$^{\arcsec}$, which we believe represents the effective length-scale of seeing-induced correlation. However the PSF for atmospheric turbulence is likely not Gaussian and is longer tailed (often described with a Moffat function). Hence, the impact of seeing is extended. This could in principle artificially increase the correlation distance across some or all time-scales. Further investigation of seeing on correlation length scales along the slit is required but is not undertaken here.

\subsection{Is there developed turbulence at 0.1~$R_\odot$ above the surface?}
The results from the power spectrum and correlation distance, taken together, would indicate that at this height in the corona, the fluctuations in the open field region are not in a state of fully developed turbulence, at least in the observed frequency range. The Doppler velocity power spectrum shows a spectral slope that is broadly in line with the expected parallel scaling from various different phenomenological theories of turbulence. Although, measurements of turbulence in the solar wind the velocity fluctuations have a scaling that is different from the magnetic spectrum, which indicates the magnetic and velocity fluctuations evolve differently towards a state of fully developed turbulence.  

Taking into account the observed relationship between parallel and perpendicular length scales, this points towards the fact that the fluctuations are not yet in a state of strong turbulence (e.g., in a regime obeying critical balance). This is almost certainly true for some fraction of the coronal Alfv\'enic waves. Some of the enhanced power around 4~mHz is thought to arise from the excitation of Alfv\'enic waves near the transition region {\citep[via mode conversion, e.g.,][]{CALGOO2008}}. These are `pristine' fluctuations that will not have experienced many turbulent non-linear interactions. Hence, time is required for the system to develop into a fully turbulent state. The rate at which the system will develop into a fully turbulent state depends on the non-linear timescale ($\tau_{NL}\sim \lambda_{corr}\ v_\perp$). Using velocity amplitudes of 20~km/s \citep[comparable to estimates from imaging data][]{MCIetal2011,MORetal2019} and perpendicular lengths scales of $\lambda_{corr}=2-8$~Mm, we obtain time-scales of $\tau_{NL}=100-400$~s. In comparison, the waves propagate relatively quickly to a height of 1.1~$R_\odot$, taking around 200-300~s \citep[assuming propagation speeds of 300-400~km/s for open field regions][]{MORetal2015}. Hence, these time-scales appear comparable and likely insufficient for the system to develop into one that is fully turbulent. It is also worth recognising that the fluctuations in open field regions are expected to be highly imbalanced, i.e., dominated by outwardly propagating waves, which can impact upon the cascade.

\subsection{Wave Energy}
Here we provide some comment on the energy of the waves but remind the reader that the amplitudes of the  Doppler velocity signal from wave fluctuations are thought to be significantly reduced by line of sight integration \citep{PASetal2011,Pant_2019,2022PhDT........10G}. Hence, we can only comment on the relative energy rather than the magnitude. The power spectrum shown in Figure~\ref{fig:high_freq} is equivalent to the kinetic energy density spectrum, $E_k(f)$ (which arises from simply multiplying the power spectrum by the plasma density). It is straightforward to infer that the high frequency waves transport less energy (power) when averaged over time than the low frequency waves (as the power spectrum essentially equates to the $v_{rms}^2$ per frequency). Not only this, but the integrated energy per logarithmic frequency interval also decreases with frequency, as the measured slope is steeper than -1. Hence the low-frequency fluctuations carry more energy. This would suggest that the energy for wave-based heating is contained in the lower frequency waves rather than high-frequency waves, which is in opposition to previous suggestions \citep[e.g.,][]{axford1992}. This result is also in apparent contradiction to recent results from Solar Orbiter which indicate high frequency waves provide the dominant contribution \citep{Lim_2023}. Although, the structures in which the high frequency waves are found are typically small bipolar regions, which are not indicative of the magnetic structure of the wider quiet Sun.  

This is generally in agreement with observations of the young solar wind from Parker Solar Probe, which indicate waves with frequencies between $3<f<20$~mHz are the predominant energy carriers \citep{Huang__2024}. Further, recent work by \cite{NAK_2025} also advocates for low frequency waves as significant energy carriers (although they analyse an event with a frequency of 0.3~mHz). As noted, while it is difficult to directly dissipate the energy in the low frequency fluctuations, (e.g., via phase mixing) turbulence could act to cascade the energy to shorter scales where dissipation mechanisms are more effective.

\section{Conclusions}
The unique capabilities of Cryo-NIRSP and DKIST have afforded us a view of the dynamics of the solar corona at short temporal scales. The high quality spectra captured
by Cryo-NIRSP enables accurate estimates of the properties of the Fe XIII coronal emission line, leading to high signal-to-noise data products. Using the shortest possible exposure times, we obtained $\sim1$~s cadence observations which are over an order of magnitude faster than previous observations readily available from the CoMP instrument.

Amongst a number of new results, we find that the Doppler velocity power spectrum extends beyond 10~mHz, maintaining a power law decline that is as at odds with expectations from models of photospheric-driven Alfv\'enic waves with strong damping in the chromosphere. Determining the source of the high-frequency fluctuations is likely not possible from observations alone and requires numerical simulations of wave propagation from the photosphere to the corona, including at least ion-neutral effects, turbulence and mode conversion. 

\section{Acknowledgments}
RJM is supported by a UKRI Future Leader Fellowship (RiPSAW—MR/T019891/1).
SRC's work was supported by the National Aeronautics and Space Administration (NASA) under grant 80NSSC20K1319, and by the National Science Foundation (NSF) under grant 2300452.

The research reported herein is based in part on data collected with the Daniel K. Inouye Solar Telescope (DKIST), a facility of the National Solar Observatory (NSO). NSO is managed by the Association of Universities for Research in Astronomy, Inc., and is funded by the National Science Foundation. DKIST is located on land of spiritual and cultural significance to Native Hawaiian people. The use of this important site to further scientific knowledge is done so with appreciation and respect.

\section{Software}
{Data analysis has been undertaken with the help of NumPy \citep{Numpy}, 
matplotlib \citep{Matplotlib}, IPython \citep{IPython}, Sunpy \citep{sunpy}, Astropy \citep{astropy}, SciPy \citep{Scipy}, PyMC \citep{pymc}.}

\bibliographystyle{aasjournal}


\appendix

\section{Uncertainties on the power spectrum}\label{app:uncert}
To fit the Bayesian model, it is useful to provide an estimate for the uncertainty on each value of power, $\sigma_{mean}(f)$. However, some care has to be taken when estimating the uncertainties on the power spectrum. The reason for this is we know that the time-series at neighbouring pixels are correlated with each other to some degree (Section~\ref{sec:coher}). Here we discuss how we estimated the uncertainties.

\medskip

When calculating the power spectrum for a single time-series, it is known that the estimate of the power at a given frequency, $\hat{S}(f)$, will be distributed about the true power, $P(f)$, following a $\chi^2_2$ distribution \citep[see, e.g.,][]{jenkinswatts,BLO,VAU2005}. Note that this only applies for a periodogram, i.e., a power spectrum estimated from the discrete Fourier transform. Upon averaging over the power spectrum of multiple ($N$) time-series, the estimate for mean value of the power at each frequency, $\mathbb{E}[P(f)]$, should follow a Normal distribution with variance $\sigma^2(f)/N$ (by the central limit theorem), where $\sigma^2(f)=\mathbb{V}[P(f)]$ is the variance of the power at frequency $f$ (which in this case is $\sigma^2(f)=P^2(f)$). For this to be be realised each estimate $\hat{S}(f)$ has to independent and identically distributed, which assumes each time-series is an independent realisation of the data generating process. However, we know from the analysis in Section~\ref{sec:coher} that the time-series have a spatial coherence, which violates the assumption of independence. 

We attempt to correct for this by dividing the number of pixels in the sample, $N$ by the estimated number of correlated pixels, $N_{cor}$, which represents the distance at which the correlation drops by $1/e$. Hence our uncertainty on the power estimate is:
$$
\sigma_{mean}(f)=\sqrt{N_{cor}(f)}\frac{\sigma_{std}(f)}{\sqrt{N}}.
$$
Figure~\ref{fig:power_spec} displays the uncertainties on power estimated in this manner and appear to give reasonable agreement with the variance observed in the average power spectrum. Although a detailed comparison suggests that the uncertainties are overestimated. For frequencies less than 1.6~mHz we use $\sigma_{mean}(f=1.6~mHz)$.

\section{Alfv\'enic wave transmission profile}\label{app:trans}

There may be some objection that the transmission profile used in Figure~\ref{fig:high_freq} is derived for torsional Alfv\'en modes and the Doppler velocities are thought to represent kink motions. However, we believe the transmission profile is reasonable. 

The merging of magnetic structures in the chromosphere and the subsequent emergence of fine-structure in the upper chromosphere (as the plasma beta drops below one) makes it difficult to envision a simplified picture of wave propagation. Mode conversion between different types of Alfv\'enic modes may be natural and unavoidable \citep[e.g.,][]{CRAVAN2005}.  For example, the kink-like motions of individual bright points \citep[e.g.,][]{NISetal2003,CHITetal2012,STAetal2013} may not be directly be associated with the kink wave motion of coronal structures. There is some evidence for mode conversion of torsional motions to kink modes in the upper chromosphere \citep{MORetal2013}. 

Moreover, the effect of ion-neutral collisions on kink waves is similar to that on torsional Alfv\'en waves \citep{SOLetal2012,SOLetal2013}. Although the rate of wave damping from Ohmic diffusion will be different. Alfv\'en waves (shear or torsional) are thought to be subject to damping via phase mixing. From the results of \cite{HEYPRI1983}, it can estimated that the propagating Alfv\'en waves have an associated damping length of
$$
L_D = \left(f^2\frac{4\pi^2\nu}{v_Al}\right)^{-1/3},
$$
where $\nu$ is the dissipative coefficient and $l$ is the length scale representative of the gradient in Alfv\'en speed.
For kink modes, the dominant damping mechanism is resonant absorption \cite{}. \cite{TERetal2010c} demonstrated for kink modes the associated damping length is
$$
L_D=f^{-1}c_{ph}\frac{2R}{\pi l}\frac{\rho_i+\rho_e}{\rho_i-\rho_e},
$$
where $c_{ph}$ is the wave propagation speed, $R$ is the radius of the flux tube, and $\rho$ is the density associated with the internal, $i$, and external, $e$ plasma of the flux tube. This indicates the rate of kink wave damping due to Ohmic diffusion is larger than that for Alfv\'en waves, hence the reduction in high frequency kink wave energy to the corona would be more pronounced that that shown by the profile in Figure~\ref{fig:high_freq}.

\section{Scaling laws based on critical balance}\label{app:scales}
The estimated power spectrum for the Doppler fluctuations indicates the existence of a power-law component of the kinetic energy spectrum. From a dimensional perspective, the power spectrum can be expressed as
\begin{equation}
  P(\omega) \, \approx \, \frac{v_{\perp}^2}{\omega}
  \, \propto \, \omega^{-\alpha} \,\,\, .
  \label{eq:Pomega}
\end{equation}
This means that the scale-dependent velocity amplitude itself
varies with frequency as $v_{\perp} \propto \omega^{(1-\alpha)/2}$.
In order for both Equations~(\ref{eq:critbal}) and
(\ref{eq:Pomega}) to be true, we must have
\begin{equation}
  k_{\perp} \, \propto \, \omega^{(1+\alpha)/2}.
  \label{eq:kperp_omega}
\end{equation}
Using Equation~(\ref{eq:kperp_omega}), the
dependence of the velocity amplitude on the perpendicular wavenumber is given by:
\begin{equation}
  v_{\perp} \, \propto \, (k_{\perp}^{2/(1+\alpha)})^{(1-\alpha)/2}
  \, \propto \, k_{\perp}^{(1-\alpha)/(1+\alpha)} \,\,\, .
  \label{eq:d3}
\end{equation}
The observed value of $\alpha = 1.58$ corresponds to
$v_{\perp} \propto k_{\perp}^{-0.225}$.
This agrees quite well with the predicted
exponent from a phenomenological theory of strong MHD turbulence
that includes scale-dependent dynamic alignment \citep[e.g.,][]{boldyrev_2006};
i.e., $v_{\perp} \sim k_{\perp}^{-1/4}$.
An {\em exact} agreement with the \citet{boldyrev_2006} value would
require a power-spectrum exponent $\alpha = 5/3$.
The above exponent differs from the theoretical exponent from 
the original critical-balance theory of \citet{Goldreich_1995} 
(i.e., $v_{\perp} \sim k_{\perp}^{-1/3}$) as well as that
predicted for weak turbulence by, e.g., \citet{Bhattacharjee_2001}
(i.e., $v_{\perp} \sim k_{\perp}^{-1/2}$). Note that Equation (\ref{eq:d3}) shows how the original critical balance theory would be consistent with $\alpha = 2$, as discussed above in Section~\ref{sec:high_freq_sub} regarding the parallel wavenumber spectrum. Our result is consistent with the growing support for the \citet{boldyrev_2006} theory that has been coming from in~situ solar wind measurements
\citep[e.g.,][]{Podesta_2009,Sioulas_2023,Sioulas_2024}, though the
in~situ data are still far from definitive.

However, we note Equation~(\ref{eq:kperp_omega}) does not seem to agree well with the observed frequency dependence of the spatial coherence of waves (Figure~\ref{fig:spatial_msc}).
If we assume the slit is perpendicular to the background magnetic
field, then the correlation distance $L_{\rm corr}$ ought to behave
essentially as the inverse of the perpendicular wavenumber,
i.e., $L_{\rm corr} \propto k_{\perp}^{-1}$.
If so, then we would expect
\begin{equation}
  L_{\rm corr} \, \propto \, \omega^{-(1+\alpha)/2}
  \label{eq:Lcorr_expected_perp}
\end{equation}
i.e., $\omega^{-1.29}$ for the observed spectrum.
This is noticeably steeper than what the coherence data seem to show.
If the other theoretical turbulent scalings discussed above
were present (i.e., with $v_{\perp} \sim k_{\perp}^{-n}$),
we would expect
\begin{equation}
  L_{\rm corr} \, \propto \, \omega^{1/(n-1)}
\end{equation}
which would be even steeper than Equation~(\ref{eq:Lcorr_expected_perp}).

\end{document}